\documentclass[12pt]{article}
\usepackage[british]{babel}
\usepackage{scalerel,stackengine}
\usepackage[utf8]{inputenc}

\def\AnswerYes{y}
\def\pdflatex{y}                  
\def\ShowLineNumberVersion{n}     
\def\ShowLabelsVersion{n}         
\def\ShowChangesVersion{n}        
\def\ShowAnnotationsVersion{n}    
\def\ShowFigures{y}               
\def\feynVersion{n}               
\def\MakeArXivLinksActive{y}      
\ifx\pdflatex\AnswerYes
   \usepackage[update,prepend]{epstopdf} 
   \usepackage{grffile} 
\fi
\ifx\pdflatex\AnswerYes
    \usepackage{color}
\else
    \usepackage[dvips]{color}
\fi
\ifx\pdflatex\AnswerYes
    \usepackage{thumbpdf}    %
\else
    \usepackage[dvips]{thumbpdf}
\fi
\ifx\pdflatex\AnswerYes
    \usepackage[ 
        final,
        breaklinks=true,
        colorlinks=true,           
        pdfborder={0 0 1},    
        citecolor={blue},linkcolor={blue},urlcolor={red},
        citebordercolor={1 0 0},linkbordercolor={0 0 1},urlbordercolor={1 0 0},
        pdfpagemode=UseOutlines,
        bookmarks=true,bookmarksopenlevel=4
        ]{hyperref}         
\else
    \usepackage[dvips,              
        final,
        breaklinks=true,
        colorlinks=true,      
        pdfborder={0 0 1},    
        citecolor={blue},linkcolor={blue},urlcolor={red},
        citebordercolor={1 0 0},linkbordercolor={0 0 1},urlbordercolor={1 0 0},
        pdfpagemode=UseOutlines,
        bookmarks=true,bookmarksopenlevel=4
         ]{hyperref}         
\fi
\ifx\MakeArXivLinksActive\AnswerYes
   \usepackage{xparse}
   \NewDocumentCommand{\arxiv} %
   {r [: u{ [} u{]]} }{[\href{http://arxiv.org/abs/#2}{arXiv:#2}~[#3]]}
   \NewDocumentCommand{\arxivold} {r[]}{[\href{http://arxiv.org/abs/#1}{#1}]}
   \NewDocumentCommand{\arXiv} %
   {r [: u{ [} u{]]} }{[\href{http://arxiv.org/abs/#2}{arXiv:#2}~[#3]]}
   \NewDocumentCommand{\arXivold} {r[]}{[\href{http://arxiv.org/abs/#1}{#1}]}
\else
   \newcommand{\arxiv}[1][]{[#1]}
   \newcommand{\arxivold}[1][]{[#1]}
   \newcommand{\arXiv}[1][]{[#1]}
   \newcommand{\arXivold}[1][]{[#1]}
\fi
\usepackage[numbers,sort&compress]{natbib}
\ifx\ShowFigures\AnswerYes
   \usepackage{graphicx}          
\else
   \renewcommand{\includegraphics}[2][]{\fbox{#2}}
\fi
\graphicspath{{figures/}{./}}
\usepackage{multirow}                   
\usepackage{amssymb}
\usepackage{amsmath}
\usepackage{bbm} 
\usepackage{xspace}                    
\xspaceaddexceptions{]}                
\usepackage{dcolumn}                    
\usepackage{bm}                        

\usepackage{cancel}	                  

\usepackage{arydshln} 
\usepackage[noaddmissingzero]{numprint} 

\usepackage{rotating}

\usepackage{floatpag}           
\floatpagestyle{plain}          
\usepackage{slashed}
\ifx\feynVersion\AnswerYes
   \usepackage{feynmp-auto} 
\fi

\usepackage{chngcntr} 
\counterwithin*{equation}{section} 

\renewcommand{\arraystretch}{1.2}

\ifx\ShowLabelsVersion\AnswerYes
   \usepackage[color]{showkeys} 
   \definecolor{refkey}{gray}{.5}   
   \definecolor{labelkey}{gray}{.5} 
   
\fi
\ifx\ShowAnnotationsVersion\AnswerYes
   \definecolor{commentblue}{rgb}{0.,0.,1.0}
   \newcommand{\comment}[1]
      {{\scriptsize\sffamily\bfseries{\textcolor{commentblue}{#1}}}}
   \newcommand{\margin}[1]{\marginpar{\scriptsize\sffamily\bfseries{#1}}}
   \newcommand{\drafty}{\textbf{Draft version \today} \hfill}
\else
   \newcommand{\comment}[1]{}
   \newcommand{\margin}[1]{}
   \newcommand{\drafty}{}
\fi
\ifx\ShowLineNumberVersion\AnswerYes
   \usepackage{lineno}
   \linenumbers
\fi
\ifx\ShowChangesVersion\AnswerYes
   \usepackage{ulem}
    
   \newcommand{\delete}[1]{\sout{#1}}            
   \renewcommand{\emph}[1]{\textit{#1}}           
\else

   \newcommand{\sout}[1]{}
   \newcommand{\xout}[1]{}

   \newcommand{\delete}[1]{}
\fi

\newcommand{\disc}{\discretionary{}{}{}}
\newcommand{\absatz}{\vspace{2ex}\noindent}

\newcommand{\cf}{\textit{cf.}\xspace}
\newcommand{\eg}{\textit{e.g.}\xspace}
\newcommand{\etal}{\textit{et al.}\xspace}
\newcommand{\etc}{\textit{etc.}\xspace}
\newcommand{\ie}{\textit{i.e.}\xspace}

\newcommand{\fs}{\scriptstyle} 

\newcommand{\hqq}{\hspace{1em}}
\newcommand{\hqqq}{\hspace{2em}}
\newcommand{\hqm}{\hspace*{-0.25em}}
\newcommand{\hqmm}{\hspace*{-0.5em}}
\newcommand{\hqmmm}{\hspace*{-1.0em}}

\newcommand{\half}{\frac{1}{2}}
\newcommand{\e}{\mathrm{e}}
\newcommand{\ii}{\mathrm{i}}
\newcommand{\dd}{\mathrm{d}}

\newcommand{\deint}[2]{\dd^{#1}\;\!\!#2\;}

\newcommand{\ev}{\vec{e}}
\newcommand{\kv}{\vec{k}}
\newcommand{\pv}{\vec{p}}
\newcommand{\qv}{\vec{q}}

\newcommand{\bra}{\langle}
\newcommand{\ket}{\rangle}

\newcommand{\mpi}{\ensuremath{m_\pi}}

\newcommand{\fpi}{\ensuremath{f_\pi}}
\newcommand{\MeV}{\ensuremath{\mathrm{MeV}}}
\newcommand{\fm}{\ensuremath{\mathrm{fm}}}
\newcommand{\ChiEFT}{$\chi$EFT\xspace}

\newcommand{\NXLO}[1]{N\ensuremath{{}^{#1}}LO\xspace}

\newcommand{\HIGS}{HI$\gamma$S\xspace}

\newcommand{\threeHe}{\ensuremath{{}^3}He\xspace}
\newcommand{\fourHe}{\ensuremath{{}^4}He\xspace}

\newcommand{\id}{\mathbbm{1}}

\newcommand{\N}{\ensuremath{\mathrm{N}}}
\newcommand{\p}{\ensuremath{\mathrm{p}}}
\newcommand{\n}{\ensuremath{\mathrm{n}}}

\newcommand{\MN}{\ensuremath{M_\mathrm{N}}} 

\newcommand{\omegacm}{\ensuremath{\omega}}
\newcommand{\thetacm}{\ensuremath{\theta}}
\newcommand{\costheta}{\ensuremath{\cos\thetacm}}

\newcommand{\jrel}{\ensuremath{j_{12}}}

\newcommand{\sepp}{,}
\newcommand{\sep}{}

\newcommand{\calO}{\mathcal{O}}

\stackMath
\renewcommand\widehat[1]{%
\savestack{\tmpbox}{\stretchto{%
  \scaleto{%
    \scalerel*[\widthof{\ensuremath{#1}}]{\kern-.6pt\bigwedge\kern-.6pt}%
    {\rule[-\textheight/2]{1ex}{\textheight}}
  }{\textheight}%
}{1.0ex}}%
\stackon[1pt]{#1}{\tmpbox}%
}

\usepackage{xstring}
\newcommand{\CG}[6]{\langle {#1} {#2}
  {\IfSubStr{#4}{+}{(#4)}{\IfSubStr{#4}{-}{(#4)}{#4}}} 
  {\IfSubStr{#5}{+}{(#5)}{\IfSubStr{#5}{-}{(#5)}{#5}}}  | 
   {#3} {\IfSubStr{#6}{+}{(#6)}{\IfSubStr{#6}{-}{(#6)}{#6}}} \rangle}

\newcommand{\wf}{}
\newcommand{\wfbra}{}
\newcommand{\sign}{(-1)}      

\newcommand{\mytitle}[1]{\begin{center}\LARGE{\textbf{#1}}\end{center}}
\newcommand{\myauthor}[1]{\textbf{#1}}
\newcommand{\myaddress}[1]{\textit{#1}}
\newcommand{\mypreprint}[1]{\begin{flushright}#1\end{flushright}}

\usepackage{tikz}

\begin{document}
%

\begin{titlepage}
  \setcounter{page}{0} \mypreprint{
    \drafty
    INT-PUB-20-022\\
    25th May 2020 \\
  }

  \mytitle{Scattering Observables from 
    One- and Two-Body Densities: Formalism and Application to $\gamma$\threeHe
    Scattering}


\begin{center}
  \myauthor{Harald W.\ Grie\3hammer$^{abc}$}\footnote{Email:
    hgrie@gwu.edu; permanent address: \emph{a}}, 
  \myauthor{Judith
    A.~McGovern$^{d}$}\footnote{Email: judith.mcgovern@manchester.ac.uk},  \\[0.5ex]
    \myauthor{Andreas Nogga$^{e}$}\footnote{Email: a.nogga@fz-juelich.de}
  \emph{and} 
  \myauthor{Daniel R.~Phillips$^{fgh}$}\footnote{Email: phillid1@ohio.edu}
  
  \vspace*{0.5cm}
  
  \myaddress{$^a$ Institute for Nuclear Studies, Department of Physics, \\The
    George Washington University, Washington DC 20052, USA}
  \\[1ex]
  \myaddress{$^b$ Department of Physics, Duke University, Box 90305, Durham NC
    27708, USA}
  \\[1ex]
  \myaddress{$^c$ High Intensity Gamma-Ray Source, Triangle Universities
    Nuclear Laboratories,\\ Box 90308, Durham NC 27708, USA}
  \\[1ex]
  \myaddress{$^d$ School of Physics and Astronomy, The University of
    Manchester,\\ Manchester M13 9PL, UK}
  \\[1ex]
  \myaddress{$^e$ IAS-4, IKP-3 and JCHP, Forschungszentrum J\"ulich, D-52428
    J\"ulich, Germany}
  \\[1ex]
  \myaddress{$^f$ Department of Physics and Astronomy and Institute of Nuclear
    and Particle Physics, Ohio University, Athens OH 45701, USA}
  \\[1ex]
  \myaddress{$^g$ Institut f\"ur Kernphysik, Technische Universit\"at
    Darmstadt, 64289 Darmstadt, Germany}
  \\[1ex]
  \myaddress{$^h$ ExtreMe Matter Institute EMMI, GSI Helmholtzzentrum f{\"u}r
    Schwerionenforschung GmbH, 64291 Darmstadt, Germany}

  \vspace*{0.2cm}

\end{center}

\vspace*{0.5cm}

\begin{abstract}
\noindent
We introduce the transition-density formalism, an efficient and general method
for calculating the interaction of external probes with light nuclei. One- and
two-body transition densities that encode the nuclear structure of the target
are evaluated once and stored. They are then convoluted with an interaction
kernel to produce amplitudes, and hence observables. By choosing different
kernels, the same densities can be used for any reaction in which a probe
interacts perturbatively with the target. The method therefore exploits the
factorisation between nuclear structure and interaction kernel that occurs in
such processes.
We study in detail the convergence in the number of partial waves for matrix
elements relevant in elastic Compton scattering on \threeHe.  The results are
fully consistent with our previous calculations in Chiral Effective Field
Theory. But the new approach is markedly more computationally efficient, which
facilitates the inclusion of more partial-wave channels in the calculation. We
also discuss the usefulness of the transition-density method for other nuclei
and reactions. Calculations of elastic Compton scattering on heavier targets
like \fourHe are straightforward extensions of this study, since the same
interaction kernels are used. And the generality of the formalism means that
our \threeHe densities can be used to evaluate any \threeHe elastic-scattering
observable with contributions from one- and two-body operators. They are
available at \url{https://datapub.fz-juelich.de/anogga}.
\end{abstract}
\vskip 1.0cm
\noindent
\begin{tabular}{rl}
  Suggested Keywords: &\begin{minipage}[t]{10.7cm} Effective Field Theory,
    Compton scattering, ab initio calculations, three-body system, few-body system,  electromagnetic reactions, reactions with external probes
                    \end{minipage}
\end{tabular}

\vskip 1.0cm

\end{titlepage}

\setcounter{footnote}{0}

\newpage

%

\section{Introduction}
\label{sec:introduction}

The structure of nuclei is usually investigated by some probe, \eg~by a
process where an external particle interacts rather weakly with the nucleons
of the nuclear target. One can then separate the dynamics of the strong
interactions that bind the nucleus from the interaction of the nucleons with
the probe and, to a high degree of accuracy, evaluate cross sections using
expectation values of well-defined operators with respect to a nuclear wave
function. For light nuclei, this has been extensively pursued for example in
electron
scattering~\cite{Golak:2005iy,Bacca:2014tla,Phillips:2016mov,Schiavilla:2018udt,Filin:2019eoe},
Compton scattering~\cite{Griesshammer:2012we}, weak decays and interactions of
neutrinos with nuclei~\cite{Baroni:2016xll,
  Engel:2016xgb,Pastore:2017uwc,Golak:2019fet} and also to investigate Physics
beyond the Standard Model~\cite{Korber:2017ery,Bsaisou:2014oka}.

Theoretical descriptions of this problem require two ingredients: first, a
reliable ``interaction kernel'', \ie.~the one- and few-body currents to which
the external probes couple (``reaction mechanism''); and second, accurate
eigenstates of the Hamiltonian for the nucleus (``structure'').  We emphasise
that this separation of the ingredients is only valid if both are evaluated
consistently in the same framework. Here we employ Chiral Effective Field
Theory ($\chi$EFT)---see \eg~refs.~\cite{Hammer:2019poc, Epelbaum:2008ga,
  Machleidt:2016rvv, Epelbaum:2019kcf} for recent reviews---as such a
framework.

In this presentation, we show that matrix elements of two-body currents can be
re-expressed as the trace of appropriately defined two-body densities with
two-body-current matrix elements. Similarly, the one-body pieces of the matrix
element are expressed as convolutions of one-body densities with the relevant
one-body operator matrix elements.  Technically, what we construct should be
called ``transition density amplitudes'' because they describe a quantum
mechanical matrix element in which the quantum numbers and momenta of the
nucleon or nucleon pair are not necessarily the same before and after the
collision. In a slight abuse of language we will, for brevity's sake, refer to
them as ``transition densities'' or ``densities''. These densities can be
directly generated from wave functions that are solutions of the
non-relativistic Schr\"odinger equation for state-of-the-art two- and
three-nucleon interactions. They are not dependent on the particular external
probe, so the nuclear-structure piece of the calculation is factorised from
the reaction mechanism.

It is hugely advantageous to strictly separate the two aspects: producing
densities on the one hand, and convoluting them with interaction kernels on
the other.

Such factorisation leads to marked gains in efficiency, since the one- and
two-body densities can be computed once and stored. This most costly part of
the evaluation can then be recycled to obtain results for a variety of
reactions on the nucleus of interest: with such densities in hand, the
evaluation of external-probe matrix elements requires only the convolution
with appropriate interaction kernels that encode the one- and two-body current
operators in the momentum-spin basis.  The computational effort associated
with this structure piece of the calculation increases significantly with $A$,
but highly parallelised and optimised codes exist that solve for the wave
functions of light nuclei.  Constructing densities from those wave functions
is straightforward.

In addition, densities can be provided in the machine-independently readable
\texttt{hdf5} format, so that other groups can use the nuclear-structure part
in a well-defined manner for their own evaluations. This separation is also
much more reliable because interaction kernels can be prepared and benchmarked
for several nuclei before applying them to previously-unstudied systems, and
because new densities can likewise be tested against known processes before
applying them to new reactions. Computational resources and development can be
focused on densities, which are then used in a variety of processes.

Indeed, separating the nuclear-structure information from the operators that
describe the interaction with external probes is not without precedent. An
analogous strategy has been used for many years in lattice QCD, where gauge
configurations for a particular lattice and lattice action are computed and
stored.  The relatively cheap evaluation of quark correlators is then carried
out separately---often for different external probes---without re-generating
the gauge configurations. Likewise, transitions using shell-model wave
functions use one-body density operators constructed from a sum of shell-model
orbitals with occupancies obtained via the diagonalisation in the model
space. Those density operators can then be contracted with a variety of
operator matrix elements to yield observables; see for example
ref.~\cite{Burrows:2017wqn,Burrows:2018ggt, Burrows:2020qvu} for recent
applications.

Here, we will go beyond simple densities and also allow for momentum transfer
into the nuclear system. This extension is facilitated by the fact that our
wave functions are obtained by solving momentum-space Faddeev equations. Thus,
in contrast to the shell-model case, the densities we employ are defined in
momentum space and are functions of the Jacobi momenta for the three-body
system, of the corresponding Faddeev angular-momentum quantum numbers, and of
momentum transfer and of energy.

We show how one- and two-body densities provide a common foundation for
elastic reactions, and illustrate their use in elastic Compton scattering as
an example of the general set-up.  This process is especially well suited for
our endeavour because a substantial fraction of its typical matrix element
comes from ``two-body currents''. The two-body densities are therefore key
elements of our approach. Of course, amplitudes involving the trace of
two-body densities play a small but important role in processes such as
electron and dark-matter scattering too, but their contribution to Compton
observables is more prominent.

In this first application, we restrict ourselves to \threeHe, whose one- and
two-body densities are publicly available
at~\url{https://datapub.fz-juelich.de/anogga}.  They are defined in momentum
space, for a wide range of both cm energies and momentum-transfers, in terms
of the Jacobi momenta for the three-body system and the corresponding Faddeev
angular-momentum quantum numbers. At present, they are based on two
combinations of local $\N\N$ and $3\N$ interactions which provide sufficiently
different, realistic numerical challenges: AV$18$ with the Urbana-IX $3\N$
interaction~\cite{Wiringa:1994wb, Pudliner:1995wk} (AV18$+$UIX), which is
relatively ``hard'' and a popular choice for testing new methods, or the
considerably softer chiral Idaho \NXLO{3} interaction at cutoff
$500\;\MeV$~\cite{Entem:2003ft} with the $\mathcal{O}(Q^3)$ \ChiEFT $3$N
interaction of variant ``b'' of ref.~\cite{Nogga:2005hp} (Idaho
\NXLO{3}{}$+3$NFb).

We check the results obtained with these densities against matrix elements
which were calculated independently using a different technique. This ensures
the numerical correctness of our ingredients and allows us to quantify the
efficiency and decrease in computational cost of the new method. Our focus is
on Compton scattering off \threeHe, where previous work obtained nuclear
matrix elements by integrating the interaction kernel directly with the wave
function of the nucleus. In the formulation of
refs.~\cite{ShuklaPhD,Choudhury:2007bh, Shukla:2008zc, Shukla:2018rzp,
  Margaryan:2018opu}, which we now call the ``traditional approach'', the
evaluation of two-body-current matrix elements for $A=3$ carried significant
numerical cost: dramatically more than for $A=2$, since extra integrations
over the momentum of a third nucleon were performed.

We also choose this process as a test since it has been the focus of several
dedicated experiments in the last decade. The High-Intensity Gamma-ray Source
(\HIGS) at the Triangle Universities Nuclear Laboratory (TUNL), the Mainz
Microtron (MAMI), and MAX-IV at Lund have all investigated elastic Compton
scattering from light nuclei including $^6$Li~\cite{Myers:2014qaa},
$^4$He~\cite{Sikora:2017rfk,Li:2019irp}, the
deuteron~\cite{Myers:2014ace,Myers:2015aba}, and others~\cite{Myers:2014qhi}.
Measurements on \threeHe are imminent at \HIGS~\cite{Ahmed} and
MAMI~\cite{Martel}.  Much of the motivation for such data is to constrain the
electromagnetic polarisabilities of the
neutron~\cite{Griesshammer:2012we,whitepaper}.

However, the computational cost of calculations in the ``traditional" approach
is prohibitive for all but the lightest targets on this list.  The new
densities-based approach opens the way for calculations of elastic Compton
scattering on \fourHe and beyond. Indeed, the extension from $A=2$ and $A=3$
to a wider range of nuclei is conceptually straightforward and does not
involve additional major computational effort---beyond that already expended
by nuclear-structure practitioners to obtain wave functions for the $A$-body
ground state. The computational cost of one- and two-body densities for, say,
the spin-zero nuclei \fourHe and $^{12}$C varies by orders of magnitude, but
the Compton convolutions use the same interaction kernels and are of
comparable computational complexity.

We emphasise that our goal here is not to provide new results for \threeHe
Compton scattering with a better description of the physics of the
process. Rather, we aim to improve the computational efficiency. The new
approach speeds up the evaluation of Compton matrix elements by a factor of
$10$ or more. This enables concomitant improvements in the numerical accuracy:
we can now include many more channels in the computation of two-body-current
matrix elements. The convergence studies presented in
sect.~\ref{sec:convergence} would come at very high computational cost in the
``traditional'' approach.

The new formulation can of course be extended in various directions.  In
processes or r\'egimes where three- and higher-body contributions to the
interaction kernel are needed, one can employ three-, four-body, \ldots,
densities, although the storage required does grow dramatically.  One can also
envision adding inelastic reactions and transmutations ($Z$ altered by
reaction), like electro-disintegration, $\beta$ decay, inelastic neutrino
scattering, inelastic Compton scattering, or photo-production of charged
pions. In this paper, however, we restrict ourselves to elastic reactions; and
hence, for Compton scattering, to energies below the pion-production
threshold, which in practice means $\omega\lesssim120\;\MeV$.

On the other hand, the present form of the framework provides an incomplete
description at low energies, since it relies on a subset of $A-2$ spectator
nucleons not participating in the reaction.  This is approximately true if the
energy inserted by the external probe is large compared to the nuclear binding
scale; see also sect.~\ref{sec:overview}. The time-scale set by the
interaction kernel is then much smaller than that of the interactions which
lead to nuclear binding. To a good approximation, the probe then interacts
with single nucleons or correlated nucleon pairs, and the nuclear response is
not collective. In Compton scattering, this is no longer the case at lower
energies~\cite{Beane:1999uq,Griesshammer:2012we}; see also
ref.~\cite{Pastore:2019}. There, ``rescattering'', namely the interaction of
all $A$ nucleons with one another between photon absorption and emission,
becomes an important reaction mechanism and should be added to the ones
calculated here. However, that is not the focus of this presentation. Rather
we are concerned with the efficient calculation of the non-collective
contributions which dominate above about 50~MeV, which is also where data is
most likely to be taken to extract nucleon polarisabilities.

\absatz The presentation is organised as follows. Section~\ref{sec:densities}
first provides an overview of the method, then explains how to construct the
one- and two-body densities for a general elastic-scattering reaction on
\threeHe, and closes with a discussion of the symmetries of those densities.
Section~\ref{sec:results} contains the premises and results of our analysis.
In sect.~\ref{sec:review}, we define the interaction kernels. The one-body
kernels are insertions of one-nucleon-spin operators with momentum
transfer. The two-nucleon kernels are those of \threeHe Compton scattering in
the variant of Chiral Effective Field Theory (\ChiEFT) with dynamical
$\Delta(1232)$ degrees of freedom at next-to-next-to-leading order [\NXLO{2},
$\mathcal{O}(e^2 \delta^3)$]~\cite{Beane:1999uq,Pascalutsa:2002pi}.  We then
discuss the amplitudes produced using one- and two-body amplitudes: choices of
$\N\N$ and $3\N$ interactions (sect.~\ref{sec:parameters}), convergence with
the number of partial waves and numerical stability
(sect.~\ref{sec:convergence}), and finally comparison to the results of the
previous, ``traditional'' approach (sect.~\ref{sec:comparison}). We
provide the customary summary and outlook in sect.~\ref{sec:conclusion}.
Appendix~\ref{sec:previous} comments on an error in the original
implementation of the one-body Compton-scattering kernel which does not affect
the results for \threeHe Compton scattering reported previously, within
expected theory uncertainties. Appendix~\ref{app:symmetries} concerns
symmetries of matrix elements.

\section{Defining Transition Densities}

\label{sec:densities}

\subsection{Overview}
\label{sec:overview}

\begin{figure}[!ht]
\begin{center}
  \includegraphics[width=\linewidth]{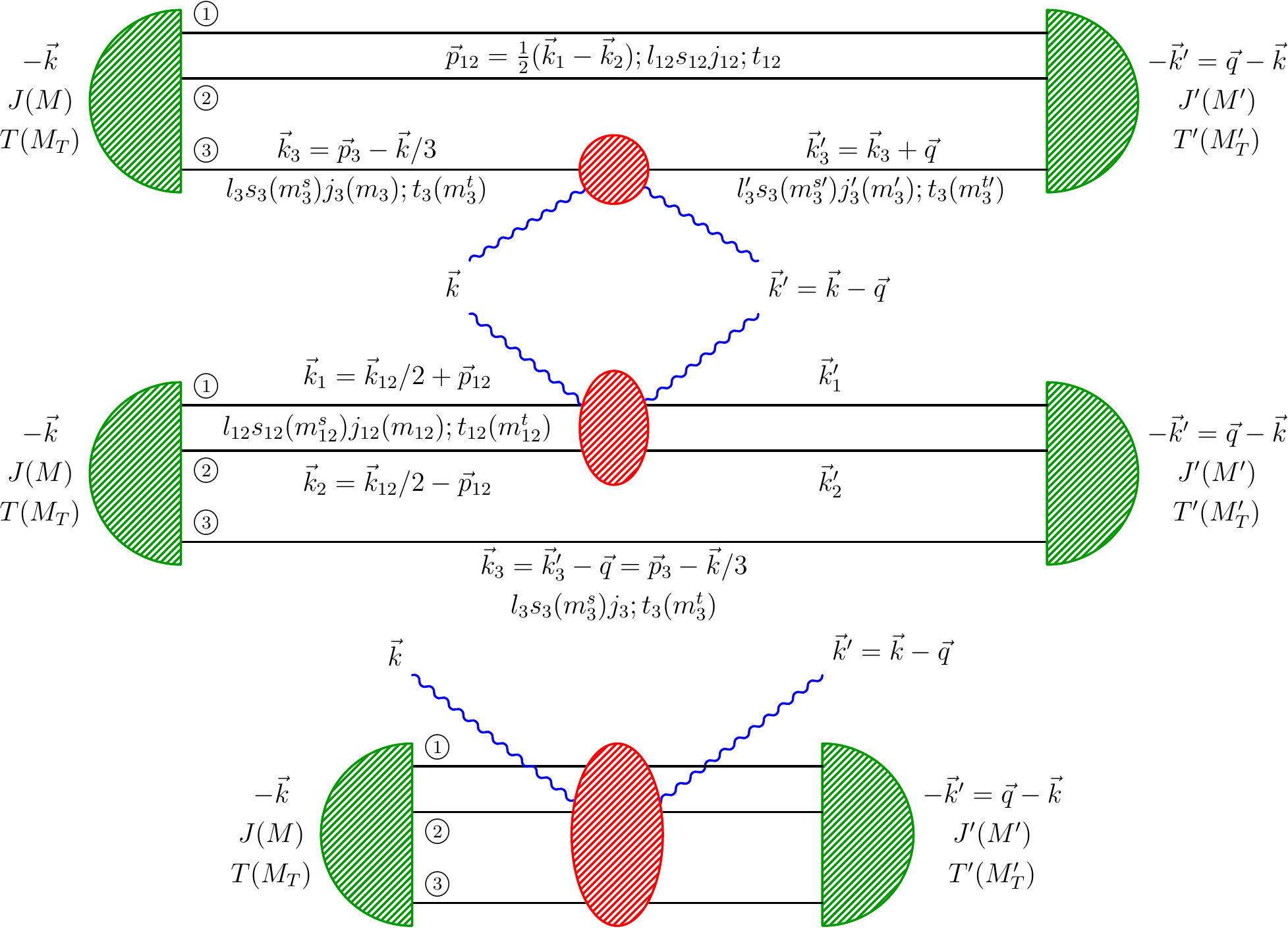}
  \caption{(Colour on-line) Kinematics for computations and convolutions
    involving transition densities of a $A=3$-nucleon bound-state, with their
    respective, pertinent quantum numbers and momentum assignments. 
    The sum over target/recoil isospins, $T/T^\prime$, is understood. While
    illustrated for Compton scattering, the assignments apply to any process
    with momentum transfer $\qv$. Top: one-body processes; centre: two-body
    processes; bottom: three-body processes (here not considered in
    detail). 
    \label{fig:kin}}
\end{center}
\end{figure}

\begin{figure}[!ht]
\begin{center}
  \includegraphics[width=0.7\linewidth]{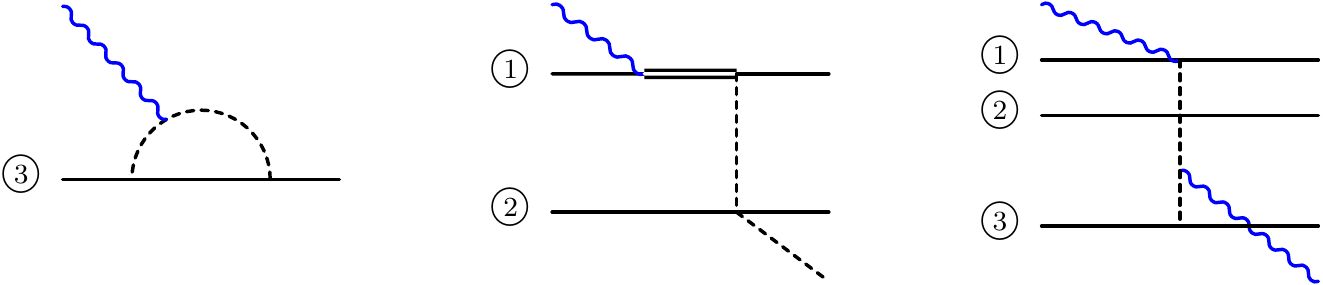}
  \caption{(Colour on-line) Illustrations of some contributions to 
    interaction kernels. Left: to the one-body kernel for form factors;
    centre: to the two-body kernel of pion-photoproduction; right: to the
    three-body kernel of Compton scattering.
    \label{fig:kernelexamples}}
\end{center}
\end{figure}

We first describe the concept, using our chosen example: Compton
scattering. As mentioned in the Introduction, one important scale is set by
the time between photon emission and photon absorption; according to the
uncertainty principle, it is $\sim 1/\omega$. If this is much smaller than the
time-scale on which nucleons interact (rescatter) inside the nucleus, then
amplitudes can be expressed as an expectation value of operators acting on the
nuclear bound state wave functions~\cite{Choudhury:2007bh, Shukla:2018rzp,
  Shukla:2008zc,ShuklaPhD}. But what is more, because the nucleus is then
``frozen in time'', the problem factorises into the Compton scattering
reaction mechanism between the photon and the $n$ \emph{active nucleons} which
directly interact with it, and a backdrop of $A-n$ \emph{spectators} which do
not\footnote{Note that for us, a ``spectator'' is every nucleon that is not
  involved in the interaction kernel. We do \emph{not} use that term for the
  ``outermost'' nucleon in Jacobi coordinates. Indeed, we choose it to be the
  spectator to a two-body matrix element, but the active participant in the
  one-body matrix element; see below.}.  This allows us to separate the
convolution into two parts: the reaction mechanism of the Compton event which
is defined by the \emph{interaction kernel} between the photon and $n$
nucleons; and the \emph{$n$-body density}. The latter is the probability
amplitude for the combination of $n$ active nucleons to change quantum
numbers, and thus accounts for the presence of all $A-n$
spectators. Figure~\ref{fig:kin} illustrates the separation, with the
interaction kernel depicted as a red ellipse. We will expand on this figure in
the subsequent presentation. Figure~\ref{fig:kernelexamples} provides example
contributions to one- and few-body interaction kernels of various reactions,
\ie~to the red ellipses of fig.~\ref{fig:kin}.

%

The \emph{one-body density} ($n=1$, illustrated for $A=3$ in the top of
fig.~\ref{fig:kin}) is the transition amplitude to find one nucleon labelled
``$A$'' with specific spin projection $m_A^s$ inside a nucleus with momentum
$-\kv$, to have it absorb a momentum transfer $\qv$ and re-arrange spin
quantum numbers, and finally be reincorporated into the $A$-body system in
such a way that the nucleus remains coherent and in its ground state.

Likewise, the \emph{two-body density} ($n=2$, centre of fig.~\ref{fig:kin} for
$A=3$) is the transition amplitude for a two-nucleon state with total momentum
$\kv_{12}$, intrinsic relative momentum of magnitude $p_{12}$ and specific
quantum numbers of the pair (relative to the $(A-2)$ spectators) to absorb a
momentum transfer $\qv$ and re-arrange its spin and angular-momentum quantum
numbers, before finally being absorbed back into the nucleus.

\emph{Few-body densities} for $n\ge3$ active nucleons can also be defined as
needed; see bottom of fig.~\ref{fig:kin} for $n=3$. Aside from the
rescattering contributions mentioned above, most important are however usually
the one-body and two-body densities.  It is a fundamental advantage of \ChiEFT
that it provides a well-defined procedure to predict such a hierarchy of
$n$-body mechanisms~\cite{Weinberg:1990rz, vanKolckPhD, vanKolck:1994yi,
  Friar:1996zw}; see also refs.~\cite{Hammer:2019poc, Epelbaum:2008ga,
  Machleidt:2016rvv, Epelbaum:2019kcf}.

We therefore carefully discuss the generation, numerical stability and
convergence of the one- and two-body densities for the $A=3$ system using
\threeHe, opening the path towards other applications of these, and of
densities for heavier nuclei like \fourHe.

\subsection{Kinematics and Partial-Wave Decomposition}
\label{sec:wavefu}

Consider a nucleus of $A$ nucleons which in the initial state has total
angular momentum $J$, spin-projection $M$ onto the $z$-axis, and
isospin-projection $M_T$ (\ie~definite charge). Several isospins $T$ may
contribute. In the \threeHe nucleus, for example, only states with total
angular momentum $J=\half$ and total isospin projection $M_T=\half$
contribute; the dominant contribution is from the iso-doublet $T=\half$, but
isospin breaking induces small $T=\frac{3}{2}$ components into the \threeHe
wave functions. A sophisticated description of nuclear processes needs to take
these into account, and ours does.

Concerning kinematics, the motion of the incident probe (the photon, in our
example) defines the $z$-axis, $\kv = |\kv|\;\ev_z$, which is also the
quantisation axis for spin-projections. Scattering takes place in the
$xz$-plane, and the momentum transfer into the nucleus is $\qv$.  We will use
the centre-of-mass (cm) frame of the probe-target system, \ie~the momentum of
the incident nucleus is $-\kv$. With our choice of conventions, three
variables suffice to characterise the process completely: the magnitude of
$\kv$ (since its direction is fixed along the $z$-axis), the magnitude of
$\qv$, and the angle between the two (since both span the scattering
plane). In our primary example, elastic Compton scattering, the outgoing
photon momentum is $\kv^{\,\prime}=\kv-\qv$ and $|\kv^{\,\prime}|=|\kv|$, so
that only two variables are independent; see sect.~\ref{sec:targetMEs}. For
electromagnetic form factors, one sets in addition $\kv=\qv$ as the momentum
of the virtual photon in the $A\gamma^\ast$ cm frame, so that there is only
one independent variable left.  The densities we produced thus far are
characterised by two independent variables only, but the formalism we discuss
now is more general.

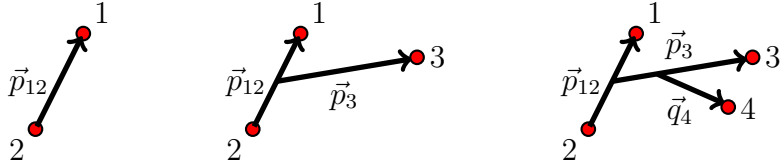
\begin{figure}[!ht]
\newcommand{\fixsize}{}
\newcommand{\fixscale}{0.6}
\begin{center}
\begin{tikzpicture}[scale=\fixscale]

\draw [fill=red,thick] (0.0-0.03,0.0-0.06) circle [radius=0.15];
\draw [fill=red,thick] (1+0.03,2+0.06) circle [radius=0.15];

\draw [->, line width=2,black] (0,0) -- (1,2);
\node[left] at (0.5,1) {\fixsize $\pv_{12}$};
\node[above right] at (1+0.03,2+0.06) {\fixsize $1$};
\node[below left] at (0.0-0.03,0.0-0.06)  {\fixsize $2$};

\end{tikzpicture}
\hspace{1cm}
\begin{tikzpicture}[scale=\fixscale]
\draw [->, line width=2,black] (0.5,1) -- (3.5,1.5);
\node[below] at (2,1.25) {\fixsize $\pv_3$};

\draw [fill=red,thick] (3.5+0.11,1.5+0.033) circle [radius=0.15];
\draw [fill=red,thick] (0.0-0.03,0.0-0.06) circle [radius=0.15];
\draw [fill=red,thick] (1+0.03,2+0.06) circle [radius=0.15];

\node[above right] at (1+0.03,2+0.06) {\fixsize $1$};
\node[below left] at (0.0-0.03,0.0-0.06)  {\fixsize $2$};
\node[right] at (3.5+0.15,1.5+0.05) {\fixsize $3$};

\draw [->, line width=2,black] (0,0) -- (1,2);
\node[left] at (0.5,1) {\fixsize $\pv_{12}$};
\end{tikzpicture}
\hspace{1cm}
\begin{tikzpicture}[scale=\fixscale]
\draw [->, line width=2,black] (1.5,1.16) -- (3,0.5);
\node[below] at (2,0.9) {\fixsize $\qv_4$};

\draw [->, line width=2,black] (0.5,1) -- (3.5,1.5);
\node[above] at (2,1.25) {\fixsize $\pv_3$};

\draw [fill=red,thick] (3.5+0.11,1.5+0.033) circle [radius=0.15];
\draw [fill=red,thick] (0.0-0.03,0.0-0.06) circle [radius=0.15];
\draw [fill=red,thick] (1+0.03,2+0.06) circle [radius=0.15];
\draw [fill=red,thick] (3+0.07,0.5-0.07) circle [radius=0.15];

\node[above right] at (1+0.03,2+0.06) {\fixsize $1$};
\node[below left] at (0.0-0.03,0.0-0.06)  {\fixsize $2$};
\node[right] at (3.5+0.15,1.5+0.05) {\fixsize $3$};
\node[right] at (3+0.1,0.5-0.1) {\fixsize $4$};

\draw [->, line width=2,black] (0,0) -- (1,2);
\node[left] at (0.5,1) {\fixsize $\pv_{12}$};
\end{tikzpicture}
\end{center}
\vspace*{-3ex}
\caption{\label{fig:jacobi} Jacobi coordinates of the two-, three- and
  four-nucleon systems, and the assignments of relative momenta $\pv,\qv$
  between constituents.}
\end{figure}
For $A\ge4$, several kinds of Jacobi momentum coordinates can be defined which
do not just differ by a permutation of the nucleons. Few-body wave functions
are however usually most efficiently represented in a hierarchical framework
like the one shown in Figure~\ref{fig:jacobi}.

Independently of the size of the nuclear systems, we always choose a
definition that singles out the $(12)$ subsystem for the application of
two-body operators, and the last ($A$th) nucleon for one-body operators,
\eg~the third one for $A=3$, or fourth for $A=4$.  As is traditional for
Jacobi coordinates, we call the $(12)$ system the ``innermost'' pair and the
$A$th nucleon the ``outermost'' nucleon. This terminology of course does not
mean that these nucleons are spatially nearest to or farthest from the centre
of the nucleus.

We now describe in detail those kinematics and quantum numbers which are
relevant for the factorised computation.  We label the momenta of the
individual nucleons in the cm frame as $\kv_1$, $\kv_2$, $\kv_3$,\dots.  The
total momentum is of course conserved and zero, $\kv+\sum_i^A\kv_i =0$. For
$A=3$, fig.~\ref{fig:kin} summarises the pertinent variables for computations
involving one-body (top) and two-body (centre) densities, as well as for
three-body densities (bottom), which are however not required for our present
purposes.

For the one-body densities, assuming that all nucleons have equal mass, the
relative momentum of the ``outermost'', active, nucleon with respect to all
others is defined by
\begin{equation}
  \pv_A = \frac{A-1}{A}\,\kv_A -\frac{1}{A}\,\sum\limits_{i=1}^{A-1}\kv_i\;\;,
\end{equation} 
see top of fig.~\ref{fig:kin}. The relative and total pair momenta of the
``innermost'' pair are
\begin{equation}
\pv_{12}=\frac{1}{2} \, \left( \kv_1 - \kv_2 \right)\;\;,\;\;
\kv_{12}=\kv_1+\kv_2\;\;.
\end{equation}
It is this ``innermost'' pair which we define as the ``active pair'' for
two-body densities, \ie~the one which interacts with the external probe; see
second row of fig.~\ref{fig:kin}.  In the $A=3$ system, which is our primary
focus, the relative momentum of the third nucleon with respect to the inner
pair is
\begin{equation}
\pv_{3}= \frac{2}{3} \, \kv_3 - \frac{1}{3} \, \left( \kv_1 + \kv_2 \right)= 
\kv_3 + \frac{1}{3}\,\kv\;\;,
\label{eq:p3}
\end{equation} 
which will allow us to later trade dependence on the momentum $\kv_3$ of an
individual nucleon for dependence on the total momentum $\kv$ of the nucleus.

We denote the \threeHe state of spin-projection $M$ by $|\wf M\ket$,
suppressing labels for the other quantum numbers $J M_T$ and bound-state
energy. This is an eigenstate of the Hamiltonian of the nucleus at rest, and
an eigenstate of both the total angular-momentum operator and its
$z$-component. We project this state onto a partial-wave-decomposed eigenstate
of the Jacobi momenta and spins (cf. below), defining a momentum-space wave
function
\begin{equation}
  \psi_{\alpha}(p_{12}p_{3}) = \bra p_{12}p_3\alpha|\wf M \ket\;\;.
  \label{eq:psialpha}
\end{equation} 
This basis is parametrised by the magnitudes of the relative pair-momentum
$p_{12}$ and the relative momentum $p_{3}$ of the third nucleon with respect
to the pair. We denote the orbital angular momentum, spin, and isospin quantum
numbers of our Jacobi-momentum basis using the collective label $\alpha$. The
orbital angular momentum $l_{12}$ and the spin $s_{12}$ of of the $(12)$
subsystem combine to give $\jrel$.  Similarly, $l_3$ and $s_3\equiv\half$
combine to give $j_3$. Finally, $\jrel$ and $j_3$ are combined into the total
angular-momentum magnitude $J$ and $z$-projection $M$ of the nucleus.  The
isospin $t_{12}$ of the $(12)$ subsystem and $t_3\equiv\half$ of the third
particle are coupled to total isospin $T$. Therefore, the quantum numbers
summarised in $\alpha$ are
\begin{equation}
    | \alpha \ket =|\left[(l_{12}s_{12})\jrel(l_3 s_3)j_3\right] JM  \sepp(t_{12}t_3)TM_T\ket\;\;.
\end{equation}
The Pauli principle guarantees that only states with
$l_{12} + s_{12} + t_{12}$ odd enter in $\alpha$.  As both $t_{12}$ and
$s_{12}$ can only have values of $0$ or $1$, the isospin of the $(12)$ pair is
actually set by
\begin{equation}
  \label{eq:t12relation}
  t_{12}=\half\;\big[1+\sign^{l_{12}+s_{12}}\big]\;\;.
\end{equation}
Since we compute $|\wf M \ket$ using isospin-violating $\N\N$ and $3\N$
interactions, the ket is \emph{not} an eigenstate of the total isospin
operator and has overlap with $\alpha$s of both $T=\half$ and $T=\frac{3}{2}$.
On the other hand, we consider \threeHe states with a specific
spin-projection, so only $\alpha$s with $J=\half$ and the appropriate $M$ have
a non-zero overlap in eq.~\eqref{eq:psialpha}.

Primed variables denote quantum numbers of the final state
$ \bra \wfbra M^\prime|$. Thus, for example, $\kv_3^{\prime}$ is the momentum
of particle $3$ when it flows into the final state, and
\begin{equation}
    \bra \alpha^\prime | =\bra\left[(l_{12}^\prime s_{12}^\prime )\jrel^\prime
      (l_3^\prime  s_3^\prime\equiv s_3)j_3^\prime \right] J^\prime M^\prime
    \sepp(t_{12}^\prime t_3^\prime\equiv t_3)T^\prime M_T^\prime |\;\;.
\end{equation}
Although we are mainly interested in $J=\half$ and $M_T=\half$ (the case of
\threeHe) we leave both arbitrary, so we can display how the formulae would
look for an arbitrary nucleus.

In this presentation, we restrict ourselves to elastic processes, \ie~the
total angular momentum $J^\prime=J$ and isospin projection $M_T^\prime=M_T$
are conserved. This implies that the probe changes neither the charge of the
struck nucleons, nor that of the spectators. Therefore, the third component of
isospin is conserved for all particles, $m^{t\prime}_i=m^t_i$ with
$i=1,\dots,A$, and for all sub-systems, \eg~$m^{t\prime}_{12}=m^t_{12}$.
However, interaction with the probe can change isospin and the wave function
of the nucleus contains components with more than one $T$, so $T^\prime\ne T$
is allowed in the densities.  Whereas fermions remain fermions
($t_3^\prime=t_3=\half$), the isospin $t_{12}$ of a fermion-pair can be
changed by interaction with the probe. The extension to include transmutation
and charge-transfer reactions ($J^\prime\ne J$ and/or $M_T^\prime\ne M_T$,
\ie~$m_i^{t\prime}\ne m_i^t$ for some nucleon(s) $i$), like charged-pion
photoproduction, $\beta$-decay or inelastic neutrino scattering, is
straightforward and left to a future publication.
 
While $J^\prime$ is identical to its unprimed counterpart, we decide to keep
its prime for out-states.  Likewise, we keep the quantum numbers $s_3=\half$
and $t_3=\half$ explicit. Both choices make it easier to track which spins and
isospins are coupled. We do however replace $m_3^{t\prime}$ by $m_3^t$, and
$m^{t\prime}_{12}$ by $m^t_{12}$.
 
Finally, we point out that the state $|\wf M\ket$ must be multiplied by an
eigenstate of the nuclear cm momentum operator, to give the momentum of the
incoming and outgoing states ($-\kv$ and $-\kv'=\qv-\kv$). Since in
non-relativistic systems, this results only in an overall momentum-conserving
$\delta^{(3)}(\kv-\kv^{\,\prime}-\qv)= \delta^{(3)}(\kv_1'+ \kv_2' + \kv_3' -
\kv_1- \kv_2 - \kv_3 - \qv)$,
we do not include these momentum wave functions explicitly in what follows.

\subsection{The One-Body Density}
\label{sec:onebodydensity}

We first consider one-body densities. We define the nucleon which interacts
with the probe to be the ``outermost'' one---the one with index $3$ in the
three-body system that we focus on; see fig.~\ref{fig:kin}. Since we represent
the nucleon as an iso-doublet consisting of the proton and neutron, the
expectation values when the photon strikes the other two nucleons are
identical and do not need to be calculated explicitly.

We start with a definition of the relevant operator and matrix element in a
basis of single-nucleon momentum, spin, and isospin states:
\begin{equation}
  \label{eq:onebody1}
  \begin{split}
 \bra \kv_{3} '  | \bra s_{3} m^{s \prime}_{3} &| \bra  t_{3} m^{t \prime}_{3} |  \hat O_{3} (\kv,\qv) | t_{3} m^{t}_{3} \ket | s_{3} m^{s}_{3} \ket |  \kv_{3}\ket \\[5pt]
  & \equiv\delta_{m^{t \prime}_{3}m^{t}_{3}}\;  \delta^{(3)}(\vec
  k_{3}' - \kv_{3}-\qv)\;  O_{3} (m^{s \prime }_{3}  m^{s}_{3}\sep
  m^{t}_{3};\kv_{3} ;\kv,\qv )  \;\;,
\end{split}
\end{equation}
where $\kv_{3}$ and $\kv_{3}'$ are the third nucleon's incoming and outgoing
momenta.  Here, the probe does not change the charge of the struck nucleon,
$m^{t \prime}_{3}=m^{t}_{3}$. The nucleon-spin components are not necessarily
conserved for spin-dependent interactions.

In this form, the probe's cm momentum $\kv$ and the momentum-transfer $\qv$
are external parameters. Momentum conservation separates off the
$\delta$-distribution in $ \kv_3^{\,\prime}-\kv_3-\qv$. For many applications,
the operators do not explicitly depend on $\kv_3$, so that the frame used for
the calculation does not matter as we will see below. But one complication
that arises in Compton scattering in $\chi$EFT at NLO and beyond is that $O_A$
explicitly depends on the single-nucleon momentum $\kv_A$. In a few- or
many-body system, this will lead to a dependence on the relative momentum
$\pv_A$ with respect to non-participating nucleons.  Such effects depend on
the nucleus and appear whenever boost corrections from the cm frame of the
nucleus to the frame of the active (struck) nucleon must be
considered. Therefore, we multipole-expand the $\kv_3$-dependence of the
spin-isospin matrix elements in spherical coordinates up to multipolarity
$K_{\max}$:
\begin{equation}
  \label{eq:opmultipole}
  \hqm O_{3} ( m_{3}^{s\, \prime} m_{3}^{s}\sep  m_{3}^{t}  ;\kv_{3} ;
  \kv,\qv ) \equiv\! \sum_{K=0}^{K_{\max}}\sum_{\kappa=-K}^{K}\hqm
  \sqrt{\frac{4\pi}{2K+1}} \left(k_{3}\right)^{K} Y_{K\kappa}(\hat k_{3})  \,
  \tilde O_{3} ( m_{3}^{s\, \prime} m_{3}^{s}\sep  m_{3}^{t}
  ;K \kappa;\kv,\qv)\;\;,
\end{equation}
where $\hat{k}$ is the unit vector (angular dependence) of $\kv$.  The
prefactors guarantee that $\tilde O_3$ and $O_3$ are identical for
$\kv_3$-independent operators or for $K=0$.  In Compton scattering, up to the
order to which we work in $\chi$EFT, it suffices to consider at most a linear
dependence of the operator on the nucleon momentum. Thus we have, so far, only
calculated one-body densities up to $K_{\max}=1$.

Let us, for concreteness, continue with \threeHe. Then, the matrix element of $\hat O_{3}$ is
written as
\begin{equation}
\begin{split}
  \bra \, \wfbra M ' \, | \, \hat O_{3} (\kv,\qv) \, | \, \wf
  \sep M \, \ket = \sum_{\alpha \alpha'}& \int \deint{}{p_{12}} p_{12}^{2}\;
  \deint{}{p_{3}} p_{3}^{2} \; \deint{}{p_{12}'} p_{12}^{\prime\, 2} \;
  \deint{}{p_{3}'} p_{3}^{\prime \, 2}\;
  \psi^{\dagger}_{\alpha'}(p_{12}'p_{3}')   \psi_{\alpha}(p_{12}p_{3})    \\[0.5ex]
  & \times \big\bra \, p_{12}' p_{3}' \left[ (l_{12}' s_{12}')\jrel'
    (l_{3}' s_3) j_{3}' \right] J' M' (t_{12}' t_3 ) T' M_{T} \, \big| \,
  \hat O_{3}(\kv,\qv) \, \\
  &\hspace*{10ex}\big| \, p_{12}p_{3} \left[ (l_{12}
    s_{12})\jrel (l_{3} s_3) j_{3} \right] J M (t_{12} t_3) T M_{T}
  \big\ket \;\; .
\end{split}
\end{equation}
Now, the matrix elements $\langle\alpha'|\hat O_{3}|\alpha\rangle$ enter, as
well as the partial-wave momentum-space wave function
$\psi_{\alpha}(p_{12}p_{3})$ of \threeHe.  Using Clebsch-Gordan coefficients
$\CG{j_{1}}{j_{2}}{j}{m_1}{m_{2}}{m}$ in the convention of
refs.~\cite{Edmonds, PDG}, we can explicitly decompose $\alpha$ so as to
separate the spin-isospin quantum numbers of the pair from those of the third
nucleon:
\begin{equation}
  \begin{split}
    \label{eq:explicitO3}
    \hqmm\bra \, &\wfbra M ' \, | \, \hat O_{3} (\kv,\qv) \, | \,
    \wf \sep M \, \ket  =\\[0.5ex]
    & \sum_{\alpha} \int \deint{}{p_{12}} p_{12}^{2} \int \deint{}{p_{3}}
    p_{3}^{2} \sum_{\alpha'} \int \deint{}{p_{12}'} p_{12}^{\prime\, 2} \int
    \deint{}{p_{3}'} p_{3}^{\prime \, 2} \psi_{\alpha'}(p_{12}'p_{3}') \;
    \psi_{\alpha}(p_{12}p_{3})    \\[0.5ex]
    &\times \sum_{m_{3}m_{3}^{t}} \CG{\jrel}{j_{3}}{J}{M-m_{3}}{m_{3}}{M}
    \CG{t_{12}}{t_3}{T}{M_{T}-m^{t}_{3}}{m^{t}_{3}}{M_{T}} \\[0.5ex]
    &\times \sum_{m_{3}'}
    \CG{\jrel'}{j_{3}'}{J'}{M'-m_{3}'}{m_{3}' }{M'}
    \CG{t_{12}'}{t_{3}}{T'}{M_{T}-m^{t}_{3}}{m^{t}_{3}}{M_{T}}
    \\[0.5ex]
    &\times \bra \, p_{12}' (l_{12}' s_{12}')\jrel' \sep (M'-m_{3}') \sepp
    t_{12}' \sep (M_{T}-m^{t}_{3}) \, | \, p_{12} (l_{12} s_{12})\jrel
    \sep (M-m_{3}) \sepp t_{12} \sep (M_{T}-m^{t}_{3}) \,
    \ket   \\[0.5ex]
    &\times \bra \, p_{3}' (l_{3}' s_3) j_{3}' \sep m_{3}' \sepp t_3 \sep
    m^{t}_{3} \, | \, \hat O_{3}(\kv,\qv) \, | \, p_{3} (l_{3}
    s_3) j_{3} \sep m_{3} \sepp t_3 \sep m^{t}_{3} \, \ket \ \ .
\end{split}
\end{equation}
Here and from now on, we directly impose the identities for spin-projections,
$M=m_3+m_{12}$, $M_T=m_3^t+m_{12}^t$ \etc The part relating to the $(12)$
subsystem of spectators just gives the usual $\delta$-distributions and a set
of Kronecker-$\delta$s in the quantum numbers of that subsystem. The matrix
element on the last line of eq.~\eqref{eq:explicitO3} also contains a
momentum-conserving $\delta$-distribution---see eq.~\eqref{eq:onebody1}. We
determine it using \eqref{eq:p3} and a corresponding relation for the primed
momenta.  This can be used to eliminate explicit dependence on $\kv_3$ in
favour of dependence on the cm momentum $\kv$ of the nucleus. The
momentum-conserving $\delta$-distribution of the operator is then
\begin{equation}
    \label{eq:delta-in-3}
  \delta^{(3)}(\kv_{3}^{\,\prime} - \kv_{3} - \qv) = \delta^{(3)}(\vec
  p_{3}^{\,\prime} - \pv_{3} - \frac{2}{3}\, \qv ) \ \ . 
\end{equation}
For the explicit evaluation of the matrix element and this
$\delta$-distribution, we insert the solid angle $\hat p_3$ of the third
particle's momentum over which we need to integrate and introduce the
three-momentum $\pv_3 = p_3 \, \hat p_3$.  The spherical harmonics then depend
on the angles of $\widehat {\pv_{3}+\frac{2}{3} \qv }$ and
$\widehat {\pv_{3}-\frac{1}{3} \kv }$ associated with linear combinations of
the three-vectors $\pv_3$, $\qv$ and $\kv$. The wave-function overlap also
needs to be evaluated at a shifted momentum magnitude
$\big| \pv_{3}+\frac{2}{3} \qv \big|$.

Inserting eqs.~\eqref{eq:onebody1} and \eqref{eq:opmultipole} into
\eqref{eq:explicitO3} gives the complete expression for the matrix element and
thus the starting point for computations in the ``traditional'' approach. What
is novel now is the recognition that the sums over quantum numbers can be
factorised. The full matrix element can thus be written as:
\begin{equation}
  \label{eq:onebodysum}
  \hqmm\hqm\bra \, \wfbra \sep M ' \, | \, \hat O_{3} (\kv,\qv) \,  | \, \wf
  \sep M \, \ket = \!\sum_{K=0}^{K_{\max}}\sum_{\kappa=-K}^{K}\! \sum_{\fs m^{s
      \prime}_{3} m^{s  }_{3}\atop\fs m_{3}^{t}}  \hqmm\tilde O_{3} (  m^{s \prime}_{3} m^{s }_{3} \sepp  m^{t}_{3} ;K \kappa
  ;\kv,\qv) \,\rho_{m_{3}^{s \prime
    }\,m_{3}^{s}}^{K\kappa;m_{3}^{t}M_{T},M'M}(\kv,\qv)\ ,
\end{equation}
where we define the one-body (transition) density by summing over those
quantum numbers that are not involved in the interaction:
\begin{equation}
  \label{eq:onebodydensity}
\begin{split}
  \rho&_{m_{3}^{s \prime }\,m_{3}^{s}}^{K\kappa;m_{3}^{t}M_{T},M'M}(\kv,\qv) :=
  \sum_{\alpha} \int \deint{}{p_{12}} p_{12}^{2} \int \deint{}{p_{3}}
  p_{3}^{2} \sum_{\alpha'} \ \delta_{\jrel\jrel'} \delta_{l_{12}l_{12}'}
  \delta_{s_{12}s_{12}'}
  \delta_{t_{12}t_{12}'}   \delta_{M_{T}M_{T}'}     \\[0.5ex]
  &\times \sum_{m_{3}} \CG{\jrel}{j_{3}'}{J'}{M-m_{3}}{M'-M+m_{3} }{M'}
  \CG{t_{12}}{t_3}{T'}{M_{T}-m^{t}_{3}}{m^{t}_{3}}{M_{T}} \\[0.5ex]
  & \hqqq \times \CG{\jrel}{j_{3}}{J}{M-m_{3}}{m_{3}}{M}
  \CG{t_{12}}{t_3}{T}{M_{T}-m^{t}_{3}}{m^{t}_{3}}{M_{T}}
  \\[0.5ex]
  & \hqqq \times \CG{l_{3}'}{s_3}{j_{3}'}{M'-M+m_{3}-m^{s \prime }_{3}}{m^{s
      \prime }_{3}}{M'-M+m_{3} }
  \CG{l_{3}}{s_3}{j_{3}}{m_{3}-m^{s}_{3}}{m^{s}_{3}}{m_{3}}   \\[0.5ex]
  & \hqqq \times\int \deint{}{\hat p_{3}} Y^{\dagger}_{l_{3}' (M'-M+m_{3}-m^{s
      \prime }_{3} ) }(\widehat {\pv_{3}+\frac{2}{3} \qv })
  \ Y_{l_{3} (m_{3}-m^{s  }_{3})  }(\hat p_{3})  \\[0.5ex]
  & \hqqq \hqqq \hqq \times \sqrt{\frac{4\pi}{2K+1}} \;\big| \vec
  p_{3}-\textstyle{\frac{1}{3}} \kv \big|^{K} \ Y_{K\kappa}(\widehat {\vec
    p_{3}-\frac{1}{3} \kv }) \;\psi^{\dagger}_{\alpha'}(p_{12} \big| \vec
  p_{3}+\textstyle{\frac{2}{3}} \qv \big| ) \ \psi_{\alpha}(p_{12}p_{3}) \;\; .
\end{split}
\end{equation}  
We present an operator form of this one-body density in
eq.~\eqref{eq:onebodyrhooperatorform}.

This convolution of operator matrix elements with the one-body density in
\eqref{eq:onebodysum} does not involve a sum over all the quantum numbers
$\{j_3l_3m^l_3m_3\}$ of the active nucleon, but only over $m^s_3$. For the
frequently-met case of $K=0$, the density also does not explicitly depend on
the momentum $-\kv$ of the nucleus. Thus it is independent of the frame chosen
and depends only on the momentum transfer $\qv$.

This density is independent of the operator $O_3$ and so allows the evaluation
of pertinent expectation values for any amplitude given by such a one-body
expectation value. The structure derived here thus applies not only to Compton
scattering, but also to many other reactions involving external probes.  Such
a separation can of course be done for any nucleus $A$, and is certainly
feasible today for \fourHe and for $p$-shell nuclei in the No-Core Shell
Model. Once a nuclear-structure calculation has been performed, the one-body
transition densities
$\rho_{m_{3}^{s \prime }\,m_{3}^{s}}^{Kk;m_{3}^{t}M_{T},M'M}(\kv,\qv)$ can be
calculated and results for processes involving external probes can be
generalised to more complex nuclei, without changing the reaction-dependent
matrix elements $\tilde O_A$. In this sense, the two parts factorise.

As an example of how to convolute one-body operators with one-body densities,
we consider the one-body contribution to a nuclear form factor. The operator
$\hat{O}_3$ must then count the number of protons or neutrons seen at a
particular momentum transfer $\qv$, \ie~at resolution $1/|\qv|$. By taking
$\qv \rightarrow 0$, such a calculation also permits us to define the
normalisation of the one-body densities.

For this $\hat{O}_3$, the matrix element $\tilde{O}_3$  in eq.~\eqref{eq:onebodysum} is
\begin{equation}
\label{eq:onebodynormop}    
\tilde{N}_3(m_3^{s'}m_3^s\sepp m_3^t;K\kappa;\kv,\qv):= 3\;
\delta_{K0}\;\delta_{\kappa0}\;\delta_{m_3^{s'}m_3^s}
\;\times\left\{
\begin{array}{ll}
\delta_{m_3^t,\half}&\mbox{for protons}\\
\delta_{m_3^t,-\half}&\mbox{for neutrons}
\end{array}\right. \;\;.
\end{equation}
Here, the symmetry factor of $3$ (counting the indistinguishable, ``active''
nucleons) must be replaced by $A$ in arbitrary nuclei. The resultant matrix
element is
\begin{equation}
    3\sum\limits_{m_3^s}\rho_{m_3^{s}\,m_3^s}^
  {(K=0)(\kappa=0);(m^t_3=\pm\half) (M_T=\half),
  M^\prime M}(\kv,\qv)\;\;,
\end{equation}
see also sect.~\ref{sec:onenucleon} for the relation to inserting Pauli spin
operators.  Summing over the single-nucleon isospins $m_3^t$ then counts the
number $A$ of nucleons. At zero momentum transfer, the matrix element should
be 2 (for $m_3^t=\frac{1}{2}$) or 1 (for $m_3^t=-\frac{1}{2}$). It follows
that the one-body density of \threeHe must obey:
\begin{equation}
\label{eq:onebodynorm}
\begin{split}
  2\delta_{M^\prime M}\stackrel{!}{=}\;& 3\sum\limits_{m_3^s}
  \;\rho_{m_3^{s}\,m_3^s}^{(K=0)(\kappa=0);(m^t_3=+\half)
    (M_T=\half),M^\prime M}(\kv,\qv=0) \;\;,
  \\
  \delta_{M^\prime M}\stackrel{!}{=}\;&3\sum\limits_{m_3^s}
  \;\rho_{m_3^{s}\,m_3^s}^{(K=0)(\kappa=0);(m^t_3=-\half) (M_T=\half),
    M^\prime M}(\kv,\qv=0)
  \;\; . 
\end{split}
\end{equation}
One-body densities are therefore dimensionless. 

Indeed, eq.~\eqref{eq:onebodydensity} shows that this normalisation is
actually imposed by normalising the \threeHe wave function to unity.  However,
the wave function is normalised by inserting a complete set of Faddeev
components $\psi_\alpha(p_{12}p_{3})$ inside \threeHe.  That partial-wave
expansion in quantum numbers $\alpha$ usually converges quickly, but is of
course truncated.  Thus, the norm of the wave function is in practice not
exactly $1$, but approaches $1$ from below, with the difference from $1$
quantifying the contribution of the missing partial waves.
Since densities are computed using a finite set of partial waves,
eq.~\eqref{eq:onebodynorm} therefore gives results which are slightly smaller
than $2$ or $1$, respectively. This is actually a better normalisation choice
when evaluating matrix elements of short- or pion-range operators since it
provides a more accurate normalisation of lower partial waves. If one were to
set the norm of the \emph{truncated} wave function to unity, contributions
from those partial waves that are included in the sum would be artificially
enhanced, distorting the results.  We will discuss the numerical deviation
from the ideal normalisation in sect.~\ref{sec:convergence1b}.

Storing one-body densities is quite cheap. For each set of kinematics $(\qv)$
and for a general spin-$J$ nucleus with isospin-projection $M_T$, there are
$2J+1$ total-spin projections $M,M^\prime$ for both the incident and outgoing
nucleons, $2$ spin-projections $m_3^s=\pm\half$ for the active nucleon both
before and after the interaction, $2$ isospin-projections
$m^t_3=m_3^{t\prime}=\pm\half$, and up to $(K_{\max}+1)^2$ entries when the
kernel needs to be multipole-expanded up to $K_{\max}$ powers of $\kv$. Thus,
not accounting for symmetries or trivial zeroes, each nucleus needs file space
for at most $8(K_{\max}+1)^2\,(2J+1)^2$ numbers per kinematic point. This is
just $128$ entries for \threeHe Compton scattering with up-to-linear boost
effects, or a few kilobytes.

\subsection{The Two-Body Density}
\label{sec:twobodydensity}

We now turn to contributions with $n=2$ ``active'' nucleons. In our
convention, these only involve quantum numbers and momenta of the two
``innermost'' nucleons, labelled $1$ and $2$; see the central diagram of
fig.~\ref{fig:kin} for the relevant kinematics and quantum numbers. Since we
treat the the nucleons in the nucleus as identical, the other pairs contribute
equally and do not need to be considered explicitly.  In Compton scattering,
this kernel parametrises the interaction of both photons with irreducible
two-nucleon currents mediated by charged pions; see fig.~\ref{fig:2Bdiagrams}.

The matrix elements of the two-nucleon operators have the form 
\begin{equation}
  \label{eq:twobody1}
  \begin{split}
    \bra \pv_{12} ' \kv_{12}^{\,\prime} |& \bra s_{12}' m^{s
      \prime}_{12} | \bra t_{12}' m^{t \prime}_{12} | \hat O_{12} (\kv,\qv) 
    | t_{12} m^{t}_{12} \ket | s_{12} m^{s}_{12} \ket | \vec
    p_{12}    \kv_{12}   \ket \\[5pt]
    & = \;\delta^{(3)}(\kv_{12}^{\,\prime} - \kv_{12}-\qv)
    \;O_{12} (
    s_{12}' t_{12}' m^{s \prime }_{12} s_{12} t_{12} m^{s}_{12} m^{t}_{12};\pv_{12}^{\, \, \prime}, \pv_{12};
   \kv,\qv )\;\;,
  \end{split}
\end{equation}
which explicitly separates out the momentum-conserving $\delta$-distribution
involving the total incoming and outgoing pair momenta $\kv_{12}$ and
$\kv_{12}'$. This operator depends on the spin $s_{12}$ and isospin $t_{12}$
of the pair, on their third components $m^{s}_{12}$ and $m^{t}_{12}$, and on
the third component $m_{12}$ of its total angular momentum $\jrel$. Besides
the photon momenta, it involves the relative momenta $\pv_{12}$ and
$\pv_{12}^{\,\prime}$ of the in- and outgoing pair. Remember that we exclude
charge-transfer or transmutation, \ie~impose $m^{t \prime}_{12}=m^{t}_{12}$
and $M_T^\prime=M_T$.

The two-nucleon operator is usually represented in terms of two-nucleon
partial-wave states with quantum numbers
\begin{equation}
  \label{eq:alpha12}
  |\alpha_{12} \ket = | (l_{12} s_{12} ) \jrel m_{12}, t_{12}m^{t
  }_{12} \ket\;\;.
\end{equation} 
If we now restrict
ourselves to the case that $O_{12}$ only depends on the momentum-transfer
$\qv$, but not on $\kv$ (see below), we write it as:
\begin{equation}
  \label{eq:2bpw}
  \begin{split}
    \bra \alpha_{12}'\kv_{12}^{\,\prime} |& \hat O_{12} |
    \alpha_{12}\kv_{12} \ket =
    \;\delta_{m^{t \prime}_{12} m^{t}_{12}}\; \delta^{(3)}(\kv_{12}^{\,\prime} -
    \kv_{12}-\qv)\\[0.5ex]
    &\times\sum_{m^{s}_{12} m^{s \prime }_{12} } \CG{l_{12}}{s_{12}}{\jrel}
    {m_{12} - m^{s}_{12}}{m^{s}_{12}}{m_{12}} \CG{l_{12}'}{s_{12}'}{\jrel'}
    {m_{12}' - m^{s \prime }_{12}}{m^{s
        \prime}_{12}}{m_{12}'}   \\[0.5ex]
    &\times \int \deint{}{\hat p_{12}'} \deint{}{\hat p_{12}} Y^{\dagger}_{l_{12}'
      (m_{12}' - m^{s \prime }_{12})} ( \hat p_{12}') \; Y_{l_{12} (m_{12} -
      m^{s }_{12})} ( \hat p_{12})
    \\[0.5ex]
    &\hqqq\times O_{12} (s_{12}' t_{12}' m^{s \prime
    }_{12} s_{12} t_{12} m^{s}_{12} m^{t}_{12} ;\pv_{12}^{\,\prime}, \pv_{12}; \qv )    \\[0.5ex]
    \equiv&\; \delta_{m^{t \prime}_{12} m^{t}_{12}} \; \delta^{(3)}(\vec
    k_{12}^{\,\prime} - \kv_{12}-\qv) \;O_{12}^{{\alpha_{12}'
        \alpha_{12}}} (p_{12}', p_{12})\;\; .
  \end{split}
\end{equation}
This must be embedded into the three-nucleon space, just like the one-body
operator. Expressed in terms of the spectator momentum, the
$\delta$-distribution of the two-nucleon operator takes the form
$ \delta^{(3)}( \pv_3 -\pv_3^{\,\prime} -\frac{1}{3} \, \qv )$.  It is then
easy to rewrite the matrix element of $\hat{O}_{12}$. As for the one-body
operator, we introduce an angular integration, evaluate the
$\delta$-distribution of the spectator's three-momentum
$\pv_3^{\,\prime} = p_3' \hat p_3 '$, and arrive at a lengthy expression used
in the ``traditional'' approach. As in the one-body case, the new method uses
the fact that its sums over quantum numbers factorise, so that we re-arrange
the matrix element as:
\begin{equation}
  \label{eq:twobodysum}
    \bra\wfbra \sep M ' \, | \hat O_{12} | \wf \sep M \, \ket \equiv
    \sum_{\alpha_{12}',\alpha_{12}} \int
    \deint{}{p_{12}} p_{12}^{2} \, \deint{}{p_{12}'} p_{12}^{\prime \, 2 } \;
    O_{12}^{\alpha_{12}' \alpha_{12}} (p_{12}', p_{12}) \;
    \rho_{\alpha_{12}'\,\alpha_{12}}^{M_{T},M'M}
    (p_{12}',p_{12};\qv)\;\;,
\end{equation}
where we define the two-body (transition) density  as:
\begin{equation}               
  \label{eq:twobodydensity}
  \begin{split}
    \rho_{\alpha_{12}'\,\alpha_{12}}^{M_{T},M'M}(p_{12}',&p_{12};\qv)
    :=\\[0.5ex]&\hspace*{-8ex}\sum_{\alpha'(\alpha_{12}')
      \alpha(\alpha_{12})} \CG{\jrel}{j_{3}}{J}{m_{12}}{M-m_{12}}{M}
    \CG{\jrel'}{j_{3}'}{J'}{m_{12}' }{M'-m_{12}'}{M'} \
    \\[0.5ex]
    & \times\CG{t_{12}}{t_3}{T}{m^{t}_{12}}{M_{T}-m^{t}_{12}}{M_{T}}
    \CG{t_{12}'}{t_3}{T'}{m^{t}_{12}}{M_{T}-m^{t}_{12}}{M_{T}} \\[0.5ex]
    & \times \int \deint{}{p_{3}'} p_{3}^{\prime \, 2 } \int \deint{}{\hat
      p_{3} '} \psi^{\dagger}_{\alpha'} (p_{12}'p_{3}') \psi_{\alpha} (p_{12}\big|
    \pv_{3} ^{{\,\prime}} +\textstyle{\frac{1}{3}}\qv
    \big| )  \\[0.5ex]
    &\hqqq\hqqq \times \sum_{m^{s}_{3}}Y^{\dagger}_{l_{3}' (M'-m_{12}' - m^{s}_{3})} (\hat p_{3}') \
    Y_{l_{3} (M-m_{12} - m^{s}_{3})} (\widehat{\pv_{3} ^{{\,\prime}}
      +\frac{1}{3}\qv }) \\[0.5ex]
    & \hqqq\hqqq\hqqq\times
    \CG{l_{3}}{s_3}{j_{3}}{M-m_{12}-m^{s}_{3}}{m^{s}_{3}}{M-m_{12}}\\[0.5ex]
    &\hqqq\hqqq\hqqq\times
    \CG{l_{3}'}{s_3}{j_{3}'}{M'-m_{12}'-m^{s}_{3}}{m^{s}_{3}}{M'-m_{12}'}\;\;.
\end{split}
\end{equation}
Here, $\alpha(\alpha_{12})$ is the set of quantum numbers $\alpha$ which
characterise the system, for a given set of quantum numbers $\alpha_{12}$ in
the $(12)$ subsystem, defined in eq.~\eqref{eq:alpha12}. As in the one-body
case, the wave functions only enter the final answer through the density.
Equations~\eqref{eq:twobodysum} and \eqref{eq:twobodydensity} make it manifest
that not all of the wave function is needed to evaluate the two-body
contribution to the matrix element; the relevant numbers are instead the
densities $\rho_{\alpha_{12}'\,\alpha_{12}}^{M_{T},M'M} (p_{12}',p_{12};\qv)$.
Once these are known for a particular wave function, evaluating the two-body
piece of the Compton amplitude is quite rapid; see discussion in
sects.~\ref{sec:convergence} and~\ref{sec:comparison}.

Note that it was not necessary to specify a particular reference frame for the
definition of the two-body density since it only depends on the momentum
transfer. This is not the case if matrix elements of $\hat{O}_{12}$ in
eq.~\eqref{eq:2bpw} also depend on $\kv_{12}$. An extension to include such
boost effects using a multipole expansion analogous to the one-body case of
eq.~\eqref{eq:opmultipole} is straightforward but not implemented in our
present file format.

We see again that the production of two-body densities and their convolution
with the two-body kernel factorise, and that the two-body densities can be
used quite generally to evaluate matrix elements involving external probes
(for now without charge-transfer reactions or boost effects).

The normalisation of the wave function requires the trace of the two-body
density at $\qv=0$ to be given by
\begin{equation}
  \label{eq:twobodynorm}
  \delta_{MM'} \stackrel{!}{=}\sum_{\alpha_{12}} \int \deint{}{p_{12}} p_{12}^2 \;
  \rho_{\alpha_{12}\,\alpha_{12}}^{M_T,M'M} (p_{12},p_{12};\qv=0)\;\;,
\end{equation}
and thus, two-body densities carry units of $\fm^3$. 

As with the one-body case in sect.~\ref{sec:onebodydensity} we can relate this
normalisation to a particular example of convoluting two-body densities and
operators. Choosing the operator
\begin{equation}
  \label{eq:twobodynormop}
  N_{12}^{{\alpha_{12}' \alpha_{12}}} (p_{12}', p_{12}) =
  3\;\delta_{l_{12}'l_{12}}\;\delta_{s_{12}'s_{12}}\;\delta_{\jrel'\jrel}\;
  \delta_{m_{12}'m_{12}}\;\frac{\delta(p_{12}'- p_{12})}{p_{12}^2}\times
  \left\{\begin{array}{ll}
      \delta_{m_{12}^t,1}&\mbox{ for $\p\p$} \\
      \delta_{m_{12}^t,0}&\mbox{ for $\n\p$} \\
      \delta_{m_{12}^t,-1}&\mbox{ for $\n\n$} 
    \end{array}\right.\;\;
\end{equation}
as $O_{12}$ in eq.~\eqref{eq:twobodysum} will yield the pair-form factor,
\ie~the number of nucleon-pairs at momentum transfer $\qv$ for a given
pair-isospin $m_{12}^t$. Here, we include a convenient factor of
${(A=3)\choose2}=3$: this is the total number of nucleon pairs in \threeHe.
Computing eq.~\eqref{eq:twobodysum} then leads to matrix elements
\begin{equation} 
  A^{M^\prime}_{M}(m_{12}^t; \qv) :=3\sum_{\fs\alpha_{12}\atop\fs \text{for
      given } m_{12}} \int
  \deint{}{p_{12}} p_{12}^2 \; 
  \rho_{\alpha_{12}\,\alpha_{12}}^{M_T,M'M} (p_{12},p_{12};\qv)
\end{equation}
which indeed count, for $\qv=0$, the number of nucleon pairs. In \threeHe:
\begin{equation} 
  \label{eq:countpairs}
  \lim\limits_{\qv\to0} A^{M^\prime}_{M}(m_{12}^t; \qv)
  \stackrel{!}{=}\delta_{M^\prime M}\times
  \left\{\begin{array}{lll}
      1&\mbox{$\p\p$ pair }  &(m_{12}^t=1)\\
      2&\mbox{$\n\p$ pairs } &(m_{12}^t=0)\\
      0&\mbox{$\n\n$ pairs } &(m_{12}^t=-1)
    \end{array}\right.\;\;.
\end{equation} 
These relations follow from eq.~\eqref{eq:twobodynorm}. In practice, the
two-body norm and eq.~\eqref{eq:countpairs} are not strictly fulfilled: the
evaluation with densities produces a result that is slightly smaller than the
correct number. This is as with the one-body normalisation in
eq.~\eqref{eq:onebodynorm} and happens for the reasons discussed there.  We
quantify this deviation from the ideal normalisation in
sect.~\ref{sec:convergence2b}.

We close this section by discussing storage for two-body densities. One- and
few-body densities all depend of course on the kinematics $\qv$ as well as the
spin and isospin of the nucleus itself (and $(K\kappa)$ if so
desired). However, the file-size of two-body densities is much larger than for
one-body densities.  First, two-body densities are subject to a much wider
range of quantum numbers of the $(12)$ sub-system. Recall from
eq.~\eqref{eq:alpha12} that
$\alpha_{12}=[(l_{12} s_{12} ) \jrel m_{12}, t_{12} m^{t }_{12}]$ and
analogously for $\alpha_{12}'$, where $s_{12}\in\{0;1\}$,
$l_{12}\in\{|\jrel-s_{12}|,\dots,\jrel+s_{12}\}$, $t_{12}\in\{0;1\}$, and
$\jrel$ goes up to a value which determines the numerical accuracy and the
convergence of the computation; see discussion in
sect.~\ref{sec:convergence2b}.  The combinations to be stored are of course
reduced by constraints on isospin and angular-momentum couplings to fit the
quantum numbers of the target, but even so, there are many more than the
corresponding eight quantum-number combinations $(m_3^{s\prime}m_3^sm_3^t)$
for a one-body density.

However, particularly costly is a dense-enough grid of momenta
$(p'_{12},p_{12})$ for the two-body density, such that the integration in
eq.~\eqref{eq:twobodysum} can be performed with sufficient accuracy; for
further discussion see sect.~\ref{sec:comparison2b}. Thus, even with very good
compression methods (\texttt{hdf5} format), two-body densities for \threeHe
reach about $10$~MB per probe energy and angle for $\jrel\le1$, about $100$~MB
for $\jrel\le2$, about $250$~MB for $\jrel\le3$, about $750$~MB for
$\jrel\le4$, and some $1,400$~MB for $\jrel\le5$. These sizes are not
exorbitant but can pose storage and memory issues in bulk computations. Files
would be reduced to half or quarter of the quoted sizes by taking advantages
of symmetry relations; see sect.~\ref{sec:symmetries}.

One should mention that, while the number of allowed quantum numbers can vary
quite a bit from nucleus to nucleus, the size of the integration grid for a
given accuracy is much less variable. That may lead to an amusing situation in
which the \emph{computation} of a two-nucleon density for the spin-$0$ nucleus
$^{12}$C is considerably more involved than that of, say, a spin-$\half$
system like $^7$Li---but the \emph{storage} needed for $^{12}$C is actually
quite a bit smaller. Computing two-body densities scales with powers of $A$,
but storing scales with powers of $J$.


\subsection{Symmetries of Few-Body Densities}
\label{sec:symmetries}

\newcommand{\phasefactor}{\ensuremath{\e^{\ii \frac{2}{3} \vec{q} \cdot
    \vec{r}_3}}}

We begin with the one-body density, which can be written in a
representation-independent operator form as:
\begin{equation}
\label{eq:onebodyrhooperatorform}
    \rho^{K \kappa; m_3^t M_T, M' M}_{m_3^{s\prime} m_3^s}(\vec{k},\vec{q})=\langle M'|\left[|s_3 m_3^{s\prime},t_3 m_3^t \rangle \;\phasefactor \;\Big[\vec{p}_3 - \tfrac{1}{3} \vec{k}\Big]^{K \kappa} \langle 
    s_3 m_3^s,t_3 m_3^t|   \right]|M 
    \rangle\;\;.
\end{equation}
Here, $\pv_3$ and $\vec{r}_3$ must be understood as operators. 
Recall from momentum conservation, eq.~\eqref{eq:delta-in-3}, that
$\frac{2}{3} \vec{q}=\pv_3^{\, \prime}-\pv_3$ is the momentum transferred to
the struck nucleon $3$ in relative coordinates. Thus, $\phasefactor$ is the
momentum-transfer operator in the nucleon cm frame.
$[\vec{a}]^{K \kappa}$ represents the $\kappa$th component of the irreducible
tensor of rank $K$ constructed from the vector $\vec{a}$; see
eq.~\eqref{eq:opmultipole}.  The projection operator in the spin-space of
particle $3$ defines that particle's spin projections.  This is combined with
a projector that enforces a specific isospin for nucleon $3$.  There is also
an implicit identity operator in the $(12)$ piece of the three-body Hilbert
space that is not written in eq.~\eqref{eq:onebodyrhooperatorform}. Inserting
complete sets of states
$|p_{12} \alpha_{12}\rangle |p_3 l_3 m_3^l\rangle |s_3 m_3^s \rangle$ and
$|p_{12} p_3 \alpha \rangle$ between the initial state and the operator so
that the wave function is evaluated in the Jacobi $\alpha$ representation and
then doing the same in the final state, verifies that
eq.~\eqref{eq:onebodyrhooperatorform} is equivalent to
eq.~\eqref{eq:onebodydensity}.

\subsubsection{Time Reversal}

As the densities are generated from strong and electromagnetic interactions,
they are time-reversal invariant. Under this symmetry, states in which angular
momenta are coupled to a total $j$ and projection $m_j$ transform
as~\cite{Edmonds}
\begin{equation}
  \label{eq:Tsym}
  \mathcal{T}  |j, m_j\ket = \sign^{j+m_j}\;|j, -m_j\ket\;\;.
\end{equation}
Likewise, when an operator $T$ is multipole-expanded into angular momentum
$(LM)$, time-reversal invariance requires (\cf~properties of spherical
harmonics)
\begin{equation}
  T_{L, -M} = \sign^{M}\; T_{LM}^\dagger\;\;.
\end{equation} 
As this symmetry will also be exploited in eq.~\eqref{eq:symmetry-twobody} for
two-body Compton matrix elements, we note that photon helicity states
transform thus as
\begin{equation}
  \label{eq:Tsym-photon}
 \mathcal{T}|\lambda\ket = \sign^\lambda \;|-\lambda\ket\;\;.
\end{equation}

These symmetries imply that one-body densities for arbitrary-spin targets
$J=J^\prime$ fulfil an exact relation when the sign of all spin-projection
quantum numbers is reversed:
\begin{equation}
  \label{eq:onebodysymmetry}
  \rho_{(-m_{3}^{s \prime })\,(-m_{3}^{s})}^{K(-\kappa);m_{3}^{t}M_{T},(-M')(-M)}(\kv,\qv) = 
  \sign^{(M^\prime-M)+(m_{3}^{s \prime}-m_{3}^{s})+\kappa}\;\rho_{m_{3}^{s
      \prime }\,m_{3}^{s}}^{K\kappa;m_{3}^{t}M_{T},M'M}(\kv,\qv)\;\;. 
\end{equation}
This cuts in half the number of one-body densities which need to be computed,
and hence the computational effort and the storage requirement---albeit the
latter is not a big deal for one-body densities.

Similarly, two-body densities obey (we do not consider multipole expansions in
$\kv_{12}$):
\begin{equation}
  \label{eq:twobodysymmetry-negate}
  \begin{split}
  \rho&_{\alpha_{12}^\prime(-m_{12}')\,\alpha_{12}(-m_{12})}^{M_{T},(-M')(-M)}
  (p_{12}',p_{12};\qv)
      \\[0.5ex]
  &\hqqq=\sign^{(M'-M)+\jrel'+\jrel+(m_{12}'-m_{12})+l_{12}'+l_{12}}\;
  \rho_{\alpha_{12}^\prime(m_{12})\,\alpha_{12}(m_{12})}^{M_{T},M'M}
  (p_{12}',p_{12};\qv) \;\;,
    \end{split}
\end{equation}
where $\alpha_{12}(-m_{12})$ is $\alpha_{12}$ of eq.~\eqref{eq:alpha12} with
the sign of $m_{12}$ reversed.

\subsubsection{Hermitecity and Parity}

Another symmetry of two-body densities guarantees that matrix elements
generated from them are Hermitean, namely the interchange of primed (outgoing)
and unprimed (incoming) quantum numbers and momenta:
\begin{equation}
  \label{eq:twobodysymmetry-revert}
  \rho_{\alpha_{12}\,\alpha_{12}^\prime}^{M_{T},MM'}
  (p_{12},p_{12}';\qv) =\sign^{l_{12}'+l_{12}}\;
  \rho_{\alpha_{12}^\prime\,\alpha_{12}}^{M_{T},M'M}
  (p_{12}',p_{12};\qv)  \;\;.
\end{equation}
To derive it, we first use the fact that the densities are real because there
are no open channels, and take the complex conjugate on both sides. Then, we
shift the integration variable in eq.~\eqref{eq:twobodydensity}, adjusting
$\pv_{3} ^{{\,\prime}}$ to $\pv_{3} ^{{\,\prime}} - \frac{\qv}{6}$, so that
the integral over $\pv_{3} ^{{\,\prime}}$ is manifestly symmetric under
$\qv \rightarrow -\qv$. After exchanging all primed and unprimed variables,
except for the integration variable $\pv_3^{\,\prime}$, the expression
contains the integral
\begin{equation}
\begin{split}
  \int\deint{}{p_{3}'} p_{3}^{\prime \, 2 } \int \deint{}{\hat
      p_{3} '}
    &\psi_{\alpha}(p_{12}\big|\pv_3^{\,\prime}
  - \textstyle{\frac{1}{6}}\qv\big|)\; \psi_{\alpha'}^\dagger(p_{12}'\big|\pv_3^{\,\prime}
  + \textstyle{\frac{1}{6}}\qv\big|)\\[0.5ex]&\times
  Y_{l_3 (M-m_{12}-m_3^s)} (\widehat{\pv_3^{\,\prime}
    - \frac{1}{6}\qv}) \;Y^\dagger_{l_3^\prime
    (M^\prime-m_{12}^\prime-m_3^s)}(\widehat{\pv_3^{\,\prime} + \frac{1}{6}\qv})\;\;.
\end{split}
\end{equation}
Changing integration variable once again,
$\pv_{3} ^{{\,\prime}} \rightarrow -\pv_{3} ^{{\,\prime}}$, and using the
parity of the spherical harmonics, $Y_{lm}(\Omega)=\sign^l\; Y_{lm}(-\Omega)$,
to reverse their arguments, reveals that this is the same integral as in
eq.~\eqref{eq:twobodydensity}, but with pre-factor $\sign^{l_3 + l_3'}$. Since
the parity of the ground state of the nucleus is unchanged by the reaction, we
use $\sign^{l_3 + l_{12}}=\sign^{l_3' + l_{12}'}$ to arrive at the
identity~\eqref{eq:twobodysymmetry-revert}.

For one-body densities, the angular integral in eq.~\eqref{eq:onebodydensity}
can likewise be rewritten by changing the integration variable
$\pv_{3} \rightarrow -\pv_{3}$ and using the parity of the spherical
harmonics. This produces the same expression, but with an additional factor
$\sign^{l_3 + l_3' + K}$ and the signs of $\vec{q}$ and $\vec{k}$ reversed. In
the one-body case, $l_{12}$ is unchanged by the interaction with the external
probe so $\sign^{l_3}=\sign^{l_3'}$. As both are integers, we arrive at the
parity relation for one-body densities:
\begin{equation}
\label{eq:onebodyparityrelation}
            \rho^{K \kappa; m_3^t M_T, M' M}_{m_3^{s\prime} m_3^s}(\vec{q},\vec{k})=\sign^{K} \rho^{K \kappa; m_3^t M_T, M' M}_{m_3^{s\prime} m_3^s}(-\vec{q},-\vec{k})\;\; .
\end{equation}
This identity is, however, of limited use, since we always choose $\vec{k}$ in
the positive $\hat{z}$ direction. It merely indicates that for even (odd) $K$,
the density is even (odd) when reversing $\qv$ and $\kv$. In particular, at
$\qv=\pm\kv$, densities must be zero for odd $K$, or their derivatives with
respect to $\qv$ and $\kv$ must be zero for even $K$.

It does tell us, though, that in the case $K=\kappa=0$ the density is even in
$\vec{q}$, and therefore real; see eq.~\eqref{eq:twobodydensity}. This leads
to a Hermitecity relation for $K=\kappa=0$ that is the one-body version of
eq.~\eqref{eq:twobodysymmetry-revert}.  It is obtained by taking the Hermitean
conjugate of eq.~\eqref{eq:onebodyrhooperatorform} to get:
\begin{equation}
            \rho^{00; m_3^t M_T, M' M}_{m_3^{s\prime} m_3^s}(\vec{q})=\rho^{00; m_3^t M_T, M M'}_{m_3^s m_3^{s\prime}}(-\vec{q})\;\; ,
\end{equation}
where we used that the density for $K=\kappa=0$ does not depend on the
initial-state momentum of the nucleus, $-\vec{k}$. Invoking
eq.~\eqref{eq:onebodyparityrelation} then produces:
\begin{equation}
\label{eq:onebodyHermitecity}
            \rho^{00; m_3^t M_T, M' M}_{m_3^{s\prime} m_3^s}(\vec{q})=\rho^{00; m_3^t M_T, M M'}_{m_3^s m_3^{s\prime}}(\vec{q})\;\; .
\end{equation}

\subsubsection{Flipping Symmetry}

\label{sec:flippingsymmetry}

So far, the symmetries discussed in this section hold for a target of
arbitrary target spin $J$. Now, we specialise to the case $J=\half$ and
elucidate an additional symmetry which relates the transition density from
states with a ``wrong-spin'' nucleon, \ie~a nucleon whose spin is anti-aligned
with the target spin, to states where the same nucleon has its spin aligned
with that of the nucleus.  The following relations then go beyond the
time-reversal relation~\eqref{eq:onebodysymmetry}:
\begin{equation}
\begin{split}
    \label{eq:onebodyflipping}
    \rho^{00; m_3^t M_T, M' M}_{(-M') M}(\vec{q})&=\rho^{00; m_3^t M_T, (-M') M}_{M' M}(\vec{q})\;\;,\\
        \rho^{00; m_3^t M_T, M' M}_{M' (-M)}(\vec{q})&=\rho^{00; m_3^t M_T, M' (-M)}_{M' M}(\vec{q}) \;\; .
        \end{split} 
\end{equation}

The presence of a wrong-spin nucleon is the key to proving them. We begin with the operator form of the density, eq.~\eqref{eq:onebodyrhooperatorform}, and choose to work in the basis 
\begin{equation}
    |p_{12} r_3 (j_{12} l_3) \ell_{123}, \mu_{123}\rangle |s_3 m_3^s \rangle\;\;.
\end{equation}
Notice that this is the only instance in this article where we employ
eigenstates of the radial coordinate $r_3$ of the $3$rd particle, and not
momentum eigenstates.  Here,
$\vec{\ell}_{123} \equiv \vec{j}_{12} + \vec{l}_3$ is the total angular
momentum of everything in the nucleus apart from the spin of the third
nucleon, so $\vec{\ell}_{123}=\vec{J}-\vec{s}_3$. Since $J=s_3=\half$ in the
case of interest, $\ell_{123}=1$ or $0$.

Let us now evaluate the density for an initial state with a ``wrong spin''
$M=-m_3^s=\pm\half$.  Since then $\mu_{123}=M-m_3^s=\pm1$, only pieces of the
initial-state ${}^3$He wave function with $\ell_{123}=1$ can
contribute. Furthermore, parity conservation ensures that the operator
$\phasefactor$ only changes $l_3$ by $0, 2, 4, \ldots$, so transitions from
$\ell_{123}=1$ to $\ell_{123}=0$ do not occur. The density can then be written
as:
\begin{equation}
\label{eq:l123basisexpression}
\begin{split}
    \rho^{00; m_3^t M_T, M' M}_{M'(-M)}&(\vec{q})=\int 
        \deint{}{p_{12}} p_{12}^2 \int \deint{}{r_3} r_3^2 \;\langle M'|\bigg[|s_3, M^\prime;m_3^t\rangle 
    |p_{12} r_3 (j_{12} l_3)1,0\rangle \\
     &\times\langle (j_{12} l_3)1,0|\;\phasefactor\;|(j_{12} l_3)1,2M \rangle 
    \; \langle (j_{12} l_3) 1,2M|\langle s_3,-M;m_3^t|\bigg]|M  \rangle \;\;.
     \end{split}
\end{equation}
That is, for transitions involving a wrong-spin nucleon, only the
$\ell_{123}=1$ piece of the wave function is needed in both the initial and
final state. We then employ the time-reversal properties of the
$|\ell_{123} \mu_{123} \rangle$ states, and of spin-half states, to derive the
matrix-element relations:
\begin{equation}
    \begin{split}
    \Big[\langle p_{12} r_3 (j_{12} l_3)1, \pm 1|\langle s_3, -M|\Big]|M\rangle&=-\Big[\langle p_{12} r_3 (j_{12} l_3) 1,\mp 1|\langle s_3, M|\Big]| -M \rangle \;\;,\\
   \langle (j_{12} l_3) 1,0|\;\phasefactor\;|(j_{12} l_3) 1, 1\rangle
   &=-\langle (j_{12} l_3) 1, 0|\;\phasefactor\;|(j_{12} l_3) 1,-1\rangle \;\;.
   \end{split}
\end{equation}
The second equality in eq.~\eqref{eq:onebodyflipping} follows immediately. The
first can be proven through the same argument, applied instead in the final
state.

We will quote the relatively straightforward generalisation of
eq.~\eqref{eq:onebodyflipping} to the case of a nucleus of arbitrary spin in a
future publication.

\subsubsection{Closing Comments on Symmetries of the Densities}

In the one-body case eq.~\eqref{eq:onebodysymmetry} reduces the number of
independent transitions densities by half, e.g., in \threeHe from $16$ to $8$
for $K=\kappa=0$ and a specific $m_3^t$. For $J=\half$ nuclei,
eq.~\eqref{eq:onebodyflipping} then relates two of the remaining $8$, with
eq.~\eqref{eq:onebodyHermitecity} removing one more. This leaves us with $5$
independent transition densities at a given $\vec{q}$ and $m_3^t$ for
$K=\kappa=0$. In fact, since eq.~\eqref{eq:l123basisexpression} expresses the
density in terms of a particular irreducible representation of a spherical
tensor, the Wigner-Eckart theorem can be used to further reduce the number of
independent one-body ($K=\kappa=0$) transition densities at a particular value
of $\vec{q}$ to three, but we do not develop those two additional relations
here.  In fact, we do not use any of the symmetry relations discussed in this
section to reduce storage requirements for the one-body density. These are
very small anyway, and keeping all $(2K+1)*16$ densities per isospin for a
given $K$ allows us to check the numerical accuracy of our results.

We finish this section with two notes. First, other variants of the symmetry
relations can be written, but they reduce to those above when one uses that
$t_{12}'+l_{12}'+s_{12}'$ and $t_{12}+l_{12}+s_{12}$ must be odd (Pauli
principle); that either both $l_{12} + l_3$ and $l_{12}' + l_3'$ are even, or
both are odd (parity of state); and that all these quantum numbers are
integers. Second, as exemplified in the argument that justifies
eq.~\eqref{eq:twobodysymmetry-revert}, in the presence of open thresholds the
symmetry relations of this section would include complex conjugation. However,
we have omitted complex conjugation in the versions we write here, instead
presenting them for the purely real densities we have computed: there are no
open channels in our calculation.

\section{Convergence and Comparisons}
\label{sec:results}

We now provide evidence that the new method speeds up the calculation of
matrix elements; that its results converge numerically; and that the converged
results agree and indeed improve both numerically and in efficiency over those
of the ``traditional'' approach. To that end, we consider in detail matrix
elements which enter in Compton scattering on \threeHe in the \ChiEFT
calculation, the specifics of which are not especially relevant in the present
context. We only use this particular process to illustrate and benchmark the
method.

\subsection{Matrix Elements for Elastic Compton Scattering}
\label{sec:review}


\subsubsection{Target Matrix Elements}
\label{sec:targetMEs}

The matrix element of the $\gamma$\threeHe amplitude depends on the spin
projections $M$ and $M^\prime$ of the incoming and outgoing nucleus onto the
$z$-axis and on the helicities $\lambda$ and $\lambda^\prime$ of the incident
and outgoing photons with polarisations $\vec{\epsilon}$ and
$\vec{\epsilon}^{\,\, \prime}$, respectively.  Using permutations, symmetries
and the same notation as in sects.~\ref{sec:onenucleon}
and~\ref{sec:twonucleon}, it is:
\begin{equation}
   \label{eq:targetme}
  A_{M\lambda}^{M^\prime\lambda^\prime}(\kv,\qv)=3\,\bra \wfbra M^\prime|
  \left[ {\hat O}^{\lambda^\prime \lambda}_3(\kv,\qv)+
    {\hat O}^{\lambda^\prime \lambda}_{12}(\kv,\qv)\right]|\wf M \ket\;\;,
\end{equation}
where the symmetry factor arises again because one of $A=3$
indistinguishable nucleons, or one of ${A=3\choose2}=3$ indistinguishable
nucleon-pairs can be struck.

The amplitude is evaluated in the cm frame of the photon-nucleus system, where
no energy is transferred. In Compton scattering, the incident-probe momentum
$\kv$ and the momentum-transfer $\qv$ which we used to characterise the
transition densities are traditionally replaced by the energy of both incident
and outgoing photon, $\omegacm=|\kv|=|\kv^{\,\prime}|$ and by the scattering
angle $\thetacm$ for the outgoing photon:
\begin{equation}
\label{eq:momtransfer}
  \costheta=1-\frac{\qv^{\,2}}{2\omegacm^2} \;\;. 
\end{equation}
From now on, we therefore discuss results using the variables
$(\omegacm,\thetacm)$ of the cm frame.

\subsubsection{One-Body Operators}
\label{sec:onenucleon}

In the cm frame of the photon-nucleon collision, the amplitude for Compton
scattering from a single nucleon is parametrised by a basis of six operators,
each of which is multiplied by an ``invariant function'' $A_i$,
$i=1, \ldots, 6$. These depend on the photon energy $\omegacm$ and scattering
angle $\thetacm=\arccos(\hat{\kv}\cdot\hat{\kv}^\prime)$, as well as the
struck nucleon's isospin. The object traditionally labelled
$T(\omegacm,\costheta)$ in the Compton literature is, in the notation for
one-body amplitudes established in sect.~\ref{sec:onebodydensity}:
\begin{equation}
  \label{eq:Tmatrix}
     \begin{split}
       O_3^{\lambda^\prime\lambda}&(m^{s\prime}_3 m^s_3\sep
       m^t_3;\kv,\qv=\kv-\kv^{\,\prime}) =\\&\bra
       m^{s\prime}_3|\bigg[A_1(\omegacm,\costheta)\;(\vec{\epsilon}\,'^\dagger\cdot
       \vec{\epsilon}) +
       A_2(\omegacm,\costheta)\;(\vec{\epsilon}\,'^\dagger\cdot\hat{k})
       \;(\vec{\epsilon} \cdot\hat{k}') \\& \hspace*{12ex}
       +\ii\,A_3(\omegacm,\costheta)\; \vec{\sigma}\cdot\left(\vec{\epsilon}\,'^\dagger\times\vec{\epsilon}\,\right)    
       +\ii\,A_4(\omegacm,\costheta)\;
       \vec{\sigma}\cdot\left(\hat{k}'\times\hat{k}\right)
       (\vec{\epsilon}\,'^\dagger\cdot\vec{\epsilon}) \\&
       \hspace*{12ex}+\ii\,A_5(\omegacm,\costheta)\;\vec{\sigma}\cdot
       \left[\left(\vec{\epsilon}\,'^\dagger\times\hat{k}
         \right)\,(\vec{\epsilon}\cdot\hat{k}')
         -\left(\vec{\epsilon}\times\hat{k}'\right)\,
         (\vec{\epsilon}\,'^\dagger\cdot\hat{k})\right] \\&
       \hspace*{12ex}+\ii\,A_6(\omegacm,\costheta)\;\vec{\sigma}\cdot
       \left[\left(\vec{\epsilon}\,'^\dagger\times\hat{k}'\right)\,
         (\vec{\epsilon}\cdot\hat{k}') -\left(\vec{\epsilon} \times\hat{k}
         \right)\, (\vec{\epsilon}\,'^\dagger\cdot\hat{k})\right]\bigg]
       |m^s_3\ket\;\;.
     \end{split}
\end{equation}
The Kronecker-$\delta$ enforces  charge conservation in electromagnetic
interactions.

Upon inspection of eq.~\eqref{eq:Tmatrix}, one infers that, for given photon
kinematics, the only independent matrix elements that actually need to be
computed to reconstruct the one-body Compton amplitude are those of the
spin-space operators which act on a single nucleon $\N$:
\begin{equation}
  \label{eq:PauliLumbanski}
  \sigma_\mu^{(\N)}:=
  (\sigma_0^{(\N)}\equiv\id^{(\N)},\sigma_x^{(\N)},\sigma_y^{(\N)},\sigma_z^{(\N)})
  \;\;.
\end{equation}  
Therefore, the analysis of sect.~\ref{sec:results} compares one-body matrix
elements with insertions of these spin operators with momentum transfer $\qv$
and no dependence on $\kv$ (\ie~multipolarity $K=\kappa=0$) between \threeHe
states:
\begin{equation}
  \label{eq:spinMEs}
  A^{M^\prime}_{M}(\sigma_\mu^{(\N)}; \qv) := 
  \bra\wfbra M^\prime|3\,\sigma_\mu^{(\N)}\,\phasefactor|\wf \sep M\ket\;\;,
\end{equation}
where we inserted again a factor of $3$ because the nucleons inside the $A=3$
nucleus are indistinguishable. For the $\mu=0$ component, this is just the
matrix element of the nucleon-number operator $N_3$ of
eq.~\eqref{eq:onebodynormop} which at $\qv=0$ yields the normalisation of the
one-body density in eq.~\eqref{eq:onebodynorm}:
\begin{equation}
  A^{M^\prime}_{M}(\sigma_0^{(\N)}; \qv) \equiv\bra\wfbra
  M^\prime|3\sigma_0^{(\N)}\,\phasefactor|\wf  M\ket\equiv
    3\sum\limits_{m_3^s}\rho_{m_3^{s}\,m_3^s}^
  {(K=0)(\kappa=0);(m^t_3=\pm\half) (M_T=\half),
  M^\prime M}(\kv,\qv)\;\;.
\end{equation}

The matrix elements of the spin insertions are not independent. They are
related to one another by rotations (and boosts) and by time reversal. The
connection between the symmetries of the densities derived in
sec.~\ref{sec:symmetries} and these relations the matrix elements
\eqref{eq:spinMEs} is discussed in app.~\ref{app:symmetries}.
  
\subsubsection{Two-Body Operators}
\label{sec:twonucleon}

The one-body case focused on the $8$ insertion operators
$\sigma_\mu^{(\N)}$. In the two-body case, we would face $16$ two-nucleon
combinations of nucleon-1-times-nucleon-2 spin-space operators
\begin{equation}
    \{\id,\vec{\sigma}_1\}\otimes\{\id,\vec{\sigma}_2\}=\id\oplus
    \vec{\sigma}_1\cdot\vec{\sigma}_2\oplus\vec{\sigma}_1\oplus\vec{\sigma}_2
    \oplus\Big[\vec{\sigma}_1\times\vec{\sigma}_2\Big]\oplus
    \Big[\sigma_1^i\sigma_2^j-
    \frac{1}{3}\delta^{ij}\vec{\sigma}_1\circ\vec{\sigma}_2\Big]
\end{equation} 
times $4$ initial- and final-state combinations of the \threeHe spin, for each
isospin $(t_{12},m_{12}^t)$ of the pair. The analogous study would provide an
overabundance of detail, in particular since it will turn out that in Compton
scattering, significant two-body contributions enter only in matrix elements
$M^\prime=M$ which do not change the \threeHe spin. Instead, we simply discuss
results for the entire ${\mathcal O}(e^2 \delta^2)$ two-body operator, \ie for
the matrix elements $A^{M^\prime,\lambda^\prime}_{M,\lambda}(\kv,\qv)$
characterised by the helicities $\lambda,\lambda^\prime$ of the incoming and
outgoing photons.

\begin{figure}[!h]
    \centering
    \includegraphics[width=0.6\linewidth]{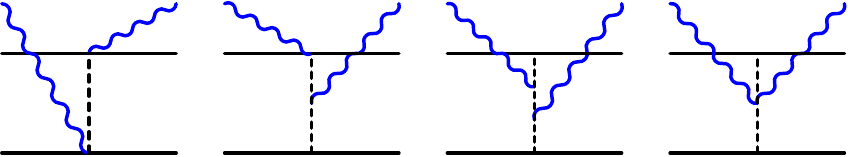}
    \caption{(Colour online) The Compton kernel for two-body contributions at
      \NXLO{2} [$\calO(e^2\delta^2)$]; crossed and permuted diagrams not
      displayed.}
    \label{fig:2Bdiagrams}
\end{figure}

In Compton scattering, the first nonzero contributions to the two-body kernel
in \ChiEFT enters at \NXLO{2} [$\calO(e^2\delta^2)$], when both photons couple
to the same charged pion-exchange current, see fig.~\ref{fig:2Bdiagrams}. They
were first computed in ref.~\cite{Beane:1999uq}, where full expressions can be
found, and were also used in all subsequent \threeHe
publications~\cite{Choudhury:2007bh, Shukla:2018rzp, Shukla:2008zc,ShuklaPhD,
  Margaryan:2018opu}.
All these two-body diagrams only contribute for $\mathrm{n}\mathrm{p}$ pairs,
\ie~they all contain an isospin factor of
$\tau^{(1)} \cdot \tau^{(2)} - \tau^{(1)}_{z} \tau^{(2)}_{z}$.  However, one
distinction between \threeHe and the deuteron is that for $A=3$, pairs with
both isospin $t_{12}=0$ and $1$ must be counted, so the summation over quantum
numbers of the $(12)$ subsystem is different.
There are no two-body corrections at $\mathcal{O}(e^2 \delta^3)$ [\NXLO{3}],
even when the $\Delta(1232)$ excitation is treated as an explicit degree of
freedom.
Finally, we list the values of the parameters of the interaction kernels in
Compton scattering on the deuteron and \threeHe~\cite{Choudhury:2007bh,
  Shukla:2018rzp, Shukla:2008zc,ShuklaPhD, Margaryan:2018opu,
  Griesshammer:2012we, Hildebrandt:2005iw}. Since the photons couple only to
the charged component of the pion-exchange currents, we use the charged-pion
mass $\mpi=139.6\;\MeV$. The pion-decay constant is $\fpi=92.42\;\MeV$; the
pion-nucleon coupling $g_A=1.267$; the fine structure constant
$\alpha_{EM}=\frac{1}{137.036}$; and $1=\hbar c=197.327\;\MeV\;\fm$.

\subsection{Choices: Interactions, Kinematic Range and Criteria}
\label{sec:parameters}

We choose two sets of $2$N and $3$N interactions to generate the one- and
two-body densities (and the wave functions for the ``traditional''
approach). The AV18 $\N\N$ model interaction~\cite{Wiringa:1994wb},
supplemented by the Urbana-IX $3\N$ interaction (3NI)~\cite{Pudliner:1995wk},
is relatively ``hard'' and thus numerically often somewhat more challenging,
while concurrently being local in coordinate space, so that a wide variety of
computational approaches can be employed. This makes it a popular choice for
testing new methods. In contradistinction, the chiral Idaho \NXLO{3}
interaction for the $2$N system at cutoff $500\;\MeV$~\cite{Entem:2003ft} with
the $\mathcal{O}(Q^3)$ \ChiEFT $3$N interaction in variant ``b'' of
ref.~\cite{Nogga:2005hp}, is also local but considerably softer. Both capture
the correct long-distance physics of one-pion exchange and reproduce both the
$\N\N$ scattering data and the experimental value of the triton and \threeHe
binding energies well.

These are of course only two out of a number of modern, sophisticated
potentials. For the purpose of this presentation, our choice is dictated by
the fact that both are already available in the ``traditional'' code, so that
they happen to be the ones most used in \threeHe Compton
scattering~\cite{Choudhury:2007bh, Shukla:2018rzp, Shukla:2008zc,ShuklaPhD,
  Margaryan:2018opu}. In sect.~\ref{sec:comparison}, we will thus also be able
to compare the results of the ``traditional'' and ``density'' approach. We
believe that they provide sufficiently different, realistic numerical
challenges.

As discussed in the Introduction and sect.~\ref{sec:overview}, we concentrate
on energies $50\;\MeV\lesssim\omegacm\lesssim120\;\MeV$, well below the
pion-production threshold, where rescattering effects are small and where
experiments are most likely to be conducted, see also detailed discussions in
refs.~\cite{Griesshammer:2012we, Margaryan:2018opu}. In this range, extremes
of energies and angles are of particular interest. At forward angles, momentum
transfers are small, and one is more sensitive to components of the densities
or wave functions which are nearly diagonal, $M^\prime=M$. Matrix elements at
back-angles (large momentum transfers) tend to be more sensitive to
off-diagonal components. Smaller energies probe long-range components, while
higher energies are more sensitive to short-distance pieces of the wave
function and the ``softness'' of the $NN$ interactions. On the other hand, one
should avoid the special symmetries which dominate the cross sections at
$0^\circ$ and $180^\circ$. We therefore illustrate our results for two extreme
(but not too extreme) choices: $(\omegacm=50\;\MeV,\thetacm=30^\circ)$, where
the momentum transfer is according to eq.~\eqref{eq:momtransfer} given by
$|\qv|=\sqrt{2\omegacm^2(1-\cos\thetacm)}=25.9\;\MeV$; and
$(\omegacm=120\;\MeV,\thetacm=165^\circ)$, where $|\qv|=237.9\;\MeV$. Results
at other energies and angles do not change our conclusions.

We use two convergence and comparison criteria: the magnitude of the relative
deviation for a given matrix element $(M^\prime M)$ at fixed energy, angle and
operator identifier $\beta$:
\begin{equation}
  \label{eq:relativedifference}
  \frac{|\Delta A^{M^\prime}_{M}(\beta;\kv,\qv)|}
  {|A^{M^\prime}_{M}(\beta;\kv,\qv)|}\;\;;
\end{equation}
and the size of the deviation of that matrix element relative to the largest
of all magnitudes of the matrix elements at the same energy and angle, over a
set of spin projections and operator identifiers:
\begin{equation}
  \label{eq:differencerelativetomax}
  \frac{|\Delta A^{M^\prime}_{M}(\beta;\kv,\qv)|}
  {\max\limits_{\{M^\prime,M;\beta\}}   
    |A^{M^\prime}_{M}(\beta;\kv,\qv)|}\;\;.
\end{equation}
For one-body matrix elements, $\beta$ is both the label $\mu$ and the particle
identifier of the spin matrix $\sigma_\mu^{(\N)}$; for two-body densities, it
is the set of photon helicities $(\lambda^\prime,\lambda)$ in the Compton
kernel.

Finally, while details of runtimes to produce one- and two-body densities are
of course highly dependent on processor and motherboard, we report them based
on our experience on the \textsc{Jureca} cluster of the J\"ulich
Supercomputing Centre (J\"ulich, Germany). The runtime magnitudes quoted for
the ``traditional'' approach and for the convolution step of the density
approach were found on a single core of a typical $7$th-generation i$7$-$4770$
with $8$ cores at $3.4$GHz. Ratios of runtimes should be fairly
processor-independent.

\subsection{Convergence of Matrix Elements in the Densities Method}
\label{sec:convergence}

The results we now present are fully converged in the radial and angular
integrations, to a relative deviation of better than $10^{-4}$. We therefore
consider now the more interesting question of convergence with respect to the
number of partial waves.

First, though, we notice that the $A=3$ system is a special case. One can
consider convergence with respect to the total angular momentum $\jrel$ of the
pair, or with respect to $j_3$, that of the ``outermost'' nucleon. The two
criteria are however not interchangeable. Specifying the quantum numbers and
momenta of the $(12)$ pair and of the nucleus as a whole determines a range of
quantum numbers of the third nucleon, but not a unique value. Imposing a
maximum $\jrel$ or $j_3$ leads therefore to differently truncated model
spaces, but both do of course converge to the same value as more partial waves
are included. We choose to examine convergence in $\jrel$ for both one- and
two-body matrix elements.

These convergence studies on \threeHe also provide experience for upcoming,
more computationally intensive computations. In \fourHe and heavier nuclei,
convergence of matrix elements is potentially more naturally discussed using
the total angular momentum of the system of ``active'' nucleons. That is still
$\jrel$ for two-body densities, and $j_{1\dots n}$ for a system of nucleons
$(1\dots n)$ which are all ``active'', \ie interact with the probe
($n>1$). For one-body matrix elements, on the other hand, one would consider
$j_A$ of the ``active'' nucleon, and not $j_{1\dots(A-1)}\to \jrel$ as we do
for $A=3$.

\subsubsection{Convergence of One-Body Matrix Elements}
\label{sec:convergence1b}

  \providecommand{\NA}{\hspace*{\fill}NA\hspace*{\fill}}
  \providecommand{\numzero}{\hspace*{\fill}{$0.$}\hspace*{\fill}}
  \renewcommand{\arraystretch}{1.2}
  \setlength{\tabcolsep}{1.7pt}
  \newcolumntype{?}[1]{!{\vrule width #1}}
  \npdecimalsign{.}
  \nprounddigits{5} 
  \npthousandthpartsep{}
  \providecommand{\stretchboth}{\rule[-1.5ex]{0pt}{4.5ex}}
  \providecommand{\stretchup}{\rule[0ex]{0pt}{3ex}}
  \providecommand{\stretchdown}{\rule[-1.5ex]{0pt}{0ex}}
  \providecommand{\linea}
     {
       {\:$\sigma_0$\!}&\stretchboth$\{\half,\half\}$}
  \providecommand{\lined}{\:$\sigma_x$\!&\stretchboth$\{\half,-\half\}$\!}
  \providecommand{\linef}{\:$\ii\sigma_y$&\stretchboth$\{\half,-\half\}$\!}
  \providecommand{\lineg}
     {\multirow{2}{*}{\:$\sigma_z$\!}&\stretchup$\{\half,\half\}$\!}
  \providecommand{\lineh}{&\stretchdown$\{\half,-\half\}$\!}
  \providecommand{\linel}{\stretchup$\{\half,\half;1,1\}$}
  \providecommand{\linem}{\stretchup$\{\half,\half;-1,1\}$}
  \providecommand{\linen}{\stretchup$\{\half,\half;1,-1\}$}
  \providecommand{\lineo}{\stretchup$\{\half,\half;-1,-1\}$}
  \providecommand{\linep}{\stretchup$\{\half,-\half;1,1\}$}
  \providecommand{\lineq}{\stretchup$\{\half,-\half;-1,1\}$}
  \providecommand{\liner}{\stretchup$\{\half,-\half;1,-1\}$}
  \providecommand{\lines}{\stretchup$\{\half,-\half;-1,-1\}$}
\begin{table}[!t]
  \small\centering
  \begin{tabular}
    {|l|l|@{\hspace{0.2ex}}
c||
   >{{\nprounddigits{4}}}n{2}{4}|
    >{{\nprounddigits{1}}}n{3}{1}:
    >{{\nprounddigits{1}}}n{3}{1}:
    >{{\nprounddigits{1}}}n{3}{1}:
    >{{\nprounddigits{1}}}n{3}{1}||
   >{{\nprounddigits{4}}}n{2}{4}|
    >{{\nprounddigits{1}}}n{3}{1}:
    >{{\nprounddigits{1}}}n{3}{1}:
    >{{\nprounddigits{1}}}n{3}{1}:
    >{{\nprounddigits{1}}}n{3}{1}|}
\multicolumn{3}{c}{}
    &\multicolumn{10}{c}{$\omegacm=50\;\MeV,\thetacm=30^\circ$}\\
    \cline{2-13}
    \multicolumn{1}{c|}{}&\multicolumn{2}{c||}{insertion}&
    \multicolumn{5}{c||}{proton}&\multicolumn{5}{c|}{neutron}\\\cline{2-13}
    \multicolumn{1}{c|}{}
    &\stretchboth\;$\sigma_\mu$\hqm&$\{M^\prime,M\}$
    &\hspace*{\fill}{value}\hspace*{\fill}
    &\hspace*{\fill}{$\fs1-\frac{\jrel\le1}{\jrel\le5}$}\hspace*{\fill}
    &\hspace*{\fill}{$\fs1-\frac{\jrel\le2}{\jrel\le5}$}\hspace*{\fill}
    &\hspace*{\fill}{$\fs1-\frac{\jrel\le3}{\jrel\le5}$}\hspace*{\fill}
    &\hspace*{\fill}{$\fs1-\frac{\jrel\le4}{\jrel\le5}$}\hspace*{\fill}
    &\hspace*{\fill}{value}\hspace*{\fill}
    &\hspace*{\fill}{$\fs1-\frac{\jrel\le1}{\jrel\le5}$}\hspace*{\fill}
    &\hspace*{\fill}{$\fs1-\frac{\jrel\le2}{\jrel\le5}$}\hspace*{\fill}
    &\hspace*{\fill}{$\fs1-\frac{\jrel\le3}{\jrel\le5}$}\hspace*{\fill}
    &\hspace*{\fill}{$\fs1-\frac{\jrel\le4}{\jrel\le5}$}\hspace*{\fill}
    \\\cline{2-13}\multicolumn{13}{c}{}\\[-3ex]\hline
    \parbox[t]{2ex}{\multirow{5}{*}{\rotatebox[origin=c]{90}{Idaho
    \NXLO{3}{}$+3$NFb}}}
&\linea
 &$1.98062$&$1.93161\%$&$0.649782\%$&$0.184351\%$&$0.0534153\%$
&$.991383$&$4.30353\%$&$0.535838\%$&$0.420774\%$&$0.0099755\%$\\
 \cline{2-13} 
 &\lined
 &$-.0414622$&$-3.68142\%$&$3.9139\%$&$-0.972634\%$&$0.371611\%$
&$.882103$&$1.32299\%$&$0.133801\%$&$0.163625\%$&$-0.0033265\%$\\
 \cline{2-13}
 &\linef
 &$-.0391802$&$-4.01894\%$&$4.06261\%$&$-1.03736\%$&$0.388025\%$
&$.882192$&$1.33209\%$&$0.133929\%$&$0.164288\%$&$-0.0034033\%$\\
 \cline{2-13}&\lineg
 &$-.0393478$&$-3.98502\%$&$4.05478\%$&$-1.03091\%$&$0.386959\%$
&$.882183$&$1.33116\%$&$0.133941\%$&$0.16413\%$&$-0.0032849\%$\\
 &\lineh
 &$.000611459$&$2.11346\%$&$1.36075\%$&$0.138616\%$&$0.0897913\%$
&$.000023851$&$91.4917\%$&$1.40294\%$&$6.73126\%$&$-0.764429\%$\\
\hline
\hline\parbox[t]{2ex}{\multirow{5}{*}{\rotatebox[origin=c]{90}{\hqmmm{}AV18$+$UIX}}}
&\linea
 &$1.97897$&$2.60512\%$&$1.1631\%$&$0.409014\%$&$0.164626\%$
&$.989348$&$5.54252\%$&$1.03426\%$&$0.860199\%$&$0.0240129\%$\\
 \cline{2-13} 
 &\lined
 &$-.0547668$&$9.15111\%$&$9.35368\%$&$1.38488\%$&$1.81292\%$
&$.861618$&$0.618982\%$&$0.0682395\%$&$0.105391\%$&$-0.0061967\%$\\
 \cline{2-13}
 &\linef
 &$-.052715$&$9.4115\%$&$9.66289\%$&$1.43204\%$&$1.88162\%$
&$.861695$&$0.627403\%$&$0.0684354\%$&$0.106179\%$&$-0.006317\%$\\
 \cline{2-13}&\lineg
 &$-.0528682$&$9.3971\%$&$9.64079\%$&$1.43053\%$&$1.87732\%$
&$.861686$&$0.626437\%$&$0.0684045\%$&$0.106012\%$&$-0.0062171\%$\\
 &\lineh
 &$.000549762$&$2.46079\%$&$1.40917\%$&$0.173218\%$&$0.04778\%$
&$.000020458$&$95.6487\%$&$2.27842\%$&$9.00342\%$&$-1.36369\%$\\
\hline
  \end{tabular}
  \npnoround
  \caption{\label{tab:convergence-onebody1} Convergence of the one-body matrix
    elements $A_M^{M\prime}(\sigma_\mu^{(\N)};\qv)$ of eq.~\eqref{eq:spinMEs} in \threeHe
    with insertions $3\sigma_\mu^{(\N)}$ for the proton and neutron, and for
    potentials Idaho \NXLO{3}{}$+3$NFb and  AV18$+$UIX in the
    ``density'' approach from $\jrel\le1$ up to $\jrel\le5$ at
    $\omegacm=50\;\MeV,\thetacm=30^\circ$, where mostly diagonal matrix
    elements $M^\prime=M$ are probed. The ``value'' column gives the results summed up to $\jrel=5$ in 
    dimensionless units, normalised as in eq.~\eqref{eq:onebodynorm}. Relative
    differences for sums to lower maximum $\jrel$, as defined in eq.~\eqref{eq:relativedifference}, are 
    shown in the subsequent columns. 
    Only those $5$ matrix elements which are independent and non-zero are
    shown. Appendix~\ref{app:symmetries} describes how time-reversal invariance relates these to the other $11$. See text for further details.}
  \end{table}
  \begin{table}[!t]
    \small\centering
    \begin{tabular}
      {|l|l|@{\hspace{0.2ex}}
      c||
      >{{\nprounddigits{4}}}n{2}{4}|
      >{{\nprounddigits{1}}}n{3}{1}:
      >{{\nprounddigits{1}}}n{3}{1}:
      >{{\nprounddigits{1}}}n{3}{1}:
      >{{\nprounddigits{1}}}n{3}{1}||
      >{{\nprounddigits{4}}}n{2}{4}|
      >{{\nprounddigits{1}}}n{3}{1}:
      >{{\nprounddigits{1}}}n{3}{1}:
      >{{\nprounddigits{1}}}n{3}{1}:
      >{{\nprounddigits{1}}}n{3}{1}|}
      \multicolumn{3}{c}{}
      &\multicolumn{10}{c}{$\omegacm=120\;\MeV,\thetacm=165^\circ$}\\
      \cline{2-13}
      \multicolumn{1}{c|}{}&\multicolumn{2}{c||}{insertion}&
                                                             \multicolumn{5}{c||}{proton}&\multicolumn{5}{c|}{neutron}\\\cline{2-13}
      \multicolumn{1}{c|}{}
      &\stretchboth$\;\sigma_\mu\hqm$&$\{M^\prime,M\}$
                                                           &\hspace*{\fill}{value}\hspace*{\fill}
                                                                                         &\hspace*{\fill}{$\fs1-\frac{\jrel\le1}{\jrel\le5}$}\hspace*{\fill}
      &\hspace*{\fill}{$\fs1-\frac{\jrel\le2}{\jrel\le5}$}\hspace*{\fill}
                               &\hspace*{\fill}{$\fs1-\frac{\jrel\le3}{\jrel\le5}$}\hspace*{\fill}
                                                           &\hspace*{\fill}{$\fs1-\frac{\jrel\le4}{\jrel\le5}$}\hspace*{\fill}
                                                                                         &\hspace*{\fill}{value}\hspace*{\fill}
      &\hspace*{\fill}{$\fs1-\frac{\jrel\le1}{\jrel\le5}$}\hspace*{\fill}
                               &\hspace*{\fill}{$\fs1-\frac{\jrel\le2}{\jrel\le5}$}\hspace*{\fill}
                                                           &\hspace*{\fill}{$\fs1-\frac{\jrel\le3}{\jrel\le5}$}\hspace*{\fill}
                                                                                         &\hspace*{\fill}{$\fs1-\frac{\jrel\le4}{\jrel\le5}$}\hspace*{\fill}
      \\\cline{2-13}\multicolumn{13}{c}{}\\[-3ex]\hline
      \parbox[t]{2ex}{\multirow{5}{*}{\rotatebox[origin=c]{90}{Idaho
      \NXLO{3}{}$+3$NFb}}}
      &\linea
                           &$.997285$&$1.81927\%$&$0.598734\%$&$0.136788\%$&$0.0375814\%$
                                                           &$.544336$&$3.68392\%$&$0.355425\%$&$0.282631\%$&$0.0023021\%$\\
      \cline{2-13} 
      &\lined
                           &$.038084$&$3.89215\%$&$-1.17058\%$&$0.775664\%$&$-0.10878\%$
                                                           &$.480055$&$1.51952\%$&$0.150754\%$&$0.183052\%$&$-0.0023874\%$\\
      \cline{2-13}
      &\linef
                           &$.0399904$&$3.79706\%$&$-1.0606\%$&$0.740249\%$&$-0.101727\%$
                                                           &$.480148$&$1.53366\%$&$0.15054\%$&$0.183644\%$&$-0.0024731\%$\\
      \cline{2-13}&\lineg
                           &$-.0699958$&$0.810715\%$&$2.39031\%$&$-0.371955\%$&$0.119174\%$
                                                           &$.474784$&$0.709441\%$&$0.163034\%$&$0.149285\%$&$0.0024074\%$\\
      &\lineh
                           &$.0144802$&$1.89747\%$&$1.13659\%$&$0.032763\%$&$0.0391593\%$
                                                           &$.000706338$&$74.4898\%$&$-0.955547\%$&$3.23596\%$&$-0.4448\%$\\
      \hline
      \hline\parbox[t]{2ex}{\multirow{5}{*}{\rotatebox[origin=c]{90}{\hqmmm{}AV18$+$UIX}}}
      &\linea
                           &$1.03766$&$2.98106\%$&$1.41758\%$&$0.470459\%$&$0.1929\%$
                                                           &$.568054$&$5.6673\%$&$1.02974\%$&$0.879516\%$&$0.0142326\%$\\
      \cline{2-13} 
      &\lined
                           &$.0220475$&$-14.291\%$&$-14.0694\%$&$-1.4713\%$&$-2.94965\%$
                                                           &$.48296$&$0.633326\%$&$0.155943\%$&$0.198421\%$&$-0.005876\%$\\
      \cline{2-13}
      &\linef
                           &$.0237721$&$-13.0906\%$&$-12.9684\%$&$-1.36267\%$&$-2.74003\%$
                                                           &$.48304$&$0.646395\%$&$0.155852\%$&$0.199301\%$&$-0.0060449\%$\\
      \cline{2-13}&\lineg
                           &$-.0757254$&$7.07276\%$&$5.5259\%$&$0.461702\%$&$0.780897\%$
                                                           &$.478447$&$-0.114665\%$&$0.161238\%$&$0.148098\%$&$0.0037695\%$\\
      &\lineh
                           &$.0130992$&$2.25522\%$&$1.10747\%$&$0.026075\%$&$-0.0601272\%$
                                                           &$.000604804$&$79.9194\%$&$-0.392267\%$&$5.53533\%$&$-1.03028\%$\\
      \hline
    \end{tabular}
    \npnoround
    \caption{\label{tab:convergence-onebody2} Convergence of the one-body matrix
      elements as in table~\ref{tab:convergence-onebody1}, but at
      $\omegacm=120\;\MeV,\thetacm=165^\circ$, where off-diagonal
      matrix elements $M^\prime\ne M$ are probed more strongly.}
\end{table}

We first turn to the results for the one-body densities in \threeHe. Here,
$\jrel$ is the relative total angular momentum between the \emph{spectators}
of the Compton process. We use a sequence $\jrel\le1,2,3,4,5$, for which
typical runtimes to produce the one-body densities should roughly increase
with the square of the number of channels $\alpha_{12}$ probed.  However,
actual runtimes scale significantly better, indicating that preparations
including I/O take most of the time. Compared to the production time for a
density with $\jrel\le1$, one with $\jrel\lesssim2$ takes only about $1.2$
times as long, compared to a factor $4$ from the rough estimate; $1.3$ times
for $\jrel\lesssim3$ (estimate: $7$ times); $1.6$ times for $\jrel\lesssim4$
(estimate: $14$ times); and $1.8$ times for $\jrel\lesssim5$ (estimate: $20$
times).
The latter takes less than $3\;\mathrm{s}$ on \textsc{Jureca}, using $8$ nodes
with $68$ cores each, or about $20$ CPU-minutes. With densities in hand,
computing the matrix elements of $\sigma_\mu^{(\N)}$, eq.~\eqref{eq:spinMEs},
in the \threeHe ground state only involves summations over quantum numbers and
is therefore nearly instantaneous. As $\sigma_y$ is imaginary in our
representation, we show the matrix elements for $\ii\sigma_y$, which are real.

As discussed in sects.~\ref{sec:onebodydensity} and \ref{sec:onenucleon}, the
matrix elements are normalised following eq.~\eqref{eq:onebodynorm} such that
an insertion of $3\sigma_0^{(\p)}$ ($3\sigma_0^{(\n)}$) at $\omegacm=0$ simply
counts the number of protons (neutrons) in the target, \ie~the result should
be $2\delta_{M^\prime M}$ ($\delta_{M^\prime M}$). These relations are
approached from below. We produced densities at zero momentum-transfer and
confirmed that at $\jrel\le5$, the relative difference is $<0.41\times10^{-3}$
for the proton and $<0.88\times10^{-3}$ for the neutron when one uses the
chiral Idaho potential, or $<1.5\times10^{-3}$ for the proton and
$<3.3\times10^{-3}$ for the neutron when one uses AV18$+$UIX. In either case,
a relative accuracy of $<7\times10^{-3}$ for the normalisation of the one-body
density is already achieved for $\jrel\le2$.

Tables~\ref{tab:convergence-onebody1} and~\ref{tab:convergence-onebody2} show
the convergence pattern and converged values of those $5$ matrix elements for
$M=\pm\half$, $M^\prime=\half$ which are nonzero and not equal to any
others. By time-reversal invariance and rotational symmetry, the other $11$
matrix elements are zero or equal to these, see app.~\ref{app:symmetries} for
further discussion. Those with $M^\prime=-\half$ follow from
eq.~\eqref{eq:Tsym} via a relation which is numerically nearly perfectly
satisfied:
\begin{equation}
  \label{eq:symmetry-onebody}
  A^{-M^\prime}_{-M}(\sigma_\mu^{(\N)};
  \qv)=\sign^{M^\prime-M+\mu}\,A^{M^\prime}_{M}(\sigma_\mu^{(\N)};
  \qv)\;\;. 
\end{equation}
Furthermore, the proofs of eqs.~\eqref{eq:sigmayzero}, \eqref{eq:sigma0zero} and \eqref{eq:sigmaxzident} in the appendix show that
\begin{equation}
    \begin{split}
        A_{\pm\half}^{\mp\half}(\sigma_0^{(\N)};\qv)=0=A_{\pm\half}^{\pm\half}(\sigma_y^{(\N)};\qv)\\
        A_{\pm\half}^{\pm\half}(\sigma_x^{(\N)};\qv)=
A_{\pm\half}^{\mp\half}(\sigma_z^{(\N)};\qv)\;\;,
    \end{split}
\end{equation}
each of which is fulfilled to better than $10^{-9}$. 

We note that the progression of the convergence is not monotonic as some wave
function components only contribute for sufficiently high $\jrel$. However,
the computations are clearly converged in $\jrel$ even at the higher energy
and momentum transfer examined in table~\ref{tab:convergence-onebody2}.

Insertions of the proton's or neutron's unit operators produce, of course, the
matrix elements with the largest magnitudes, of order $1$. For the small
momentum transfer at $(\omega=50\;\MeV, \theta=30^\circ)$, these matrix
elements deviate only a little from the zero-energy normalisation of
eq.~\eqref{eq:onebodynorm}. At the highest momentum transfer,
$(\omega=120\;\MeV, \theta=165^\circ)$, their values have dropped, as it
happens to about a factor of one-half. Similarly large are the neutron's
spin-operators $\sigma_x^{(\n)}$ and $\sigma_y^{(\n)}$ for
$(M^\prime=\half,M=-\half)$, and $\sigma_z^{(\n)}$ for
$(M^\prime=\half,M=\half)$. This is well-explained by a na\"ive model of
\threeHe as two protons paired to total spin zero, plus a neutron which
therefore carries all the \threeHe spin. This class of matrix elements is
converged in $\jrel$ to $\lesssim0.2\%$. It shows little dependence on the
potential, typically $\lesssim2\%$ for $(\omegacm=50\;\MeV,\thetacm=30^\circ)$
and $\lesssim4\%$ at the higher momentum transfer,
$(\omegacm=120\;\MeV,\thetacm=165^\circ)$.

A second class consists of those matrix elements which are suppressed by
factors of $10$ to $100$: insertions of the proton's spin-operators
$\sigma_z^{(\p)}$ and $\sigma_x^{(\p)}$ for any combination $(M^\prime,M)$, or
of $\sigma_y^{(\p)}$ for $(M^\prime=\half,M=-\half)$. For each potential
separately, they are converged to better than $3\%$. The difference between
results for the two potentials can be as large as $30\%$, but we note that
this translates to only $\lesssim0.5\%$ relative to the magnitude of the
largest matrix elements as defined in
eq.~\eqref{eq:differencerelativetomax}. Not surprisingly, this is a signal
that these matrix elements are more sensitive to short-range details of the
two- and three-nucleon interactions. In the na\"ive \threeHe model of paired
proton-spins and only $s$-wave interactions, they are zero. For realistic
interactions, they are nonzero because of the small $d$-wave and
$s^\prime$-wave contributions in which the proton spins are parallel. Most
modern potentials yield a $d$-wave ``probability'' of $\lesssim10\%$ in
\threeHe, and while this is of course not an observable, it is instructive as
an indicator of how complex a potential is to solve numerically. AV18$+$UIX,
with its hard core and correspondingly higher amount of two-body $sd$-mixing
creates larger matrix elements which also converge more slowly.

All the other matrix elements are $\lesssim10^{-3}$ and thus usually
irrelevant for observables. With differences of $\lesssim10\%$ between
potentials, they are again more sensitive to short-distance details. But they
are converged for each individual potential to between $\lesssim1\%$ and
$\lesssim0.001\%$ relative to the magnitude of the largest matrix elements,
\cf~\eqref{eq:differencerelativetomax}.

We take convergence in $\jrel$ as a sign of numerical stability, and
differences between potentials as signs of some residual theoretical
uncertainties associated with short-distance pieces of the Compton response
that are of higher order in \ChiEFT. Therefore, we find that relative to the
magnitude of the largest one-body matrix elements defined in
eq.~\eqref{eq:differencerelativetomax}, the numerical error never exceeds
$0.2\%$. Theoretical uncertainties due to potential choice are less than $2\%$
at the smaller momentum transfer, or $4\%$ at the larger one. There are, of
course, other theoretical uncertainties induced by the truncation of the
\ChiEFT Compton kernels. Now, we are not concerned with those, but refer to
their assessment in ref.~\cite{Margaryan:2018opu}.

\subsubsection{Convergence of Two-Body Matrix Elements}
\label{sec:convergence2b}

Turning now to the results for two-body densities, we consider the matrix
elements when the two-body kernel coupling the two Compton photons to the
charged exchange-pion is inserted; see sect.~\ref{sec:twonucleon}.  We use a
sequence $\jrel\le1,2,3,4$ to study convergence. Like for the
one-body-densities, production times for two-body densities should roughly
increase with the square of the number of channels, but they must be produced
for a sufficiently fine grid of momenta $(p^\prime_{12},\p_{12})$.  Actual
runtimes scale somewhat better for small $\jrel$, indicating that preparations
are relatively less time-consuming. Compared to the production time for a
density with $\jrel\le1$, one with $\jrel\lesssim2$ takes about $2.4$ times as
long, compared to a factor $4$ from the rough estimate; $6$ times for
$\jrel\lesssim3$ (estimate: $7$ times); $16$ times for $\jrel\lesssim4$
(estimate: $14$ times); and $29$ times for $\jrel\lesssim5$ (estimate: $20$
times).
Computing a two-body density with $\jrel\le5$ takes about fifteen times longer
than the corresponding one-body density, namely about $35\;\mathrm{s}$ on
\textsc{Jureca}, using $8$ nodes with $68$ cores each, or less than $6$
CPU-hours.  In order to go from densities to matrix elements, one must perform
the radial and angular integrations over $\pv_{12}$ and summation over quantum
numbers in the $(12)$ subsystem, see eq.~\eqref{eq:twobodysum}. On a
workstation, that adds less than half an hour per energy and angle for
$\jrel\le1$, a few hours or a factor of $\approx8$ for $\jrel\le2$, a workday
or another factor of $\approx3$ for $\jrel\le3$, a full day or another factor
of $\approx3$ for $\jrel\le4$, and two full days or another factor of $2$ for
$\jrel\le5$.  One could expedite this by multipole-expanding the kernel, but
the same computations in the ``traditional'' approach take many times that;
see sect.~\ref{sec:comparison2b}.

As we did for the one-body densities, we discuss the extent to which the
normalisation of eq.~\eqref{eq:twobodynorm} is fulfilled. We confirmed that
for densities at zero momentum-transfer and $\jrel\le4$, the relative
difference is $\lesssim5.2\times10^{-3}$.  Individually, the operator of
eq.~\eqref{eq:countpairs} undercounts both the number of proton and neutron
pairs by about the same $0.5\%$, each. In either case, a relative accuracy of
$<6.6\times10^{-3}$ for the normalisation of the two-body density is already
achieved for $\jrel\le2$.

Tables~\ref{tab:convergence-twobody1} and~\ref{tab:convergence-twobody2} show
convergence patterns and converged matrix elements for $M^\prime=\half$,
normalised following eq.~\eqref{eq:twobodynorm} and quoted in $\fm^3$.  Matrix
elements with $M^\prime=-\half$ follow again from time-reversal invariance
(see app.~\ref{app:symmetries}), with eqs.~\eqref{eq:Tsym} and
\eqref{eq:Tsym-photon} translating to:
\begin{equation}
  \label{eq:symmetry-twobody}
    A^{-M^\prime,-\lambda^\prime}_{-M,-\lambda}=
    \sign^{M^\prime-M+\lambda^\prime-\lambda}\,A^{M^\prime,\lambda^\prime}_{M,\lambda}\;\;.
\end{equation}
Matrix elements with $M^\prime\ne M$ are not displayed since they have
magnitudes $<2\times10^{-7}\;\fm^3$, \ie~are smaller than those with
$M^\prime=M$ by a factor of $<10^{-4}$. They have no impact on observables but
are quite susceptible to numerical noise. The pair of matrix elements with
$(\lambda=\lambda^\prime)$ appear to be identical in the table, as do those
with $(\lambda=-\lambda^\prime)$. Closer inspection reveals small relative
differences of magnitude $\le10^{-6}$, attributed to numerical noise.

\begin{table}[!t]
  \small\centering
  \begin{tabular}{|c|
   >{{\nprounddigits{5}}}n{2}{5}|
    >{{\nprounddigits{2}}}n{3}{2}:
    >{{\nprounddigits{2}}}n{3}{2}:
    >{{\nprounddigits{2}}}n{3}{2}||
   >{{\nprounddigits{5}}}n{2}{5}|
    >{{\nprounddigits{2}}}n{3}{2}:
    >{{\nprounddigits{2}}}n{3}{2}:
    >{{\nprounddigits{2}}}n{3}{2}|}
\multicolumn{1}{c}{}
    &\multicolumn{8}{c}{$\omegacm=50\;\MeV,\thetacm=30^\circ$}\\
    \cline{2-9}
    \multicolumn{1}{c|}{}&
    \multicolumn{4}{c||}{Idaho \NXLO{3}{}$+3$NFb}
    &\multicolumn{4}{c|}{AV18$+$UIX}\\\hline
    \stretchboth $\,\{M^\prime,M;\lambda^\prime,\lambda\}$
    &\hspace*{\fill}{value [$\fm^3$]}\hspace*{\fill}
    &\hspace*{\fill}{$\fs1-\frac{\jrel\le1}{\jrel\le4}$}\hspace*{\fill}
    &\hspace*{\fill}{$\fs1-\frac{\jrel\le2}{\jrel\le4}$}\hspace*{\fill}
    &\hspace*{\fill}{$\fs1-\frac{\jrel\le3}{\jrel\le4}$}\hspace*{\fill}
    &\hspace*{\fill}{value [$\fm^3$]}\hspace*{\fill}
    &\hspace*{\fill}{$\fs1-\frac{\jrel\le1}{\jrel\le4}$}\hspace*{\fill}
    &\hspace*{\fill}{$\fs1-\frac{\jrel\le2}{\jrel\le4}$}\hspace*{\fill}
    &\hspace*{\fill}{$\fs1-\frac{\jrel\le3}{\jrel\le4}$}\hspace*{\fill}
    \\\hline\hline
\linel
 &$-.071393$&$0.64628\%$&$0.0972\%$&$0.02914\%$
&$-.0939448$&$1.1511\%$&$0.54308\%$&$0.05854\%$\\
 \linem
 &$-.0054336$&$0.67726\%$&$0.11042\%$&$0.03312\%$
&$-.0070626$&$1.15256\%$&$0.54654\%$&$0.06514\%$\\
 \linen
 &$-.0054336$&$0.67726\%$&$0.11042\%$&$0.03312\%$
&$-.0070624$&$1.14976\%$&$0.54656\%$&$0.06514\%$\\
 \lineo
 &$-.071393$&$0.64628\%$&$0.0972\%$&$0.02942\%$
&$-.0939448$&$1.1511\%$&$0.54308\%$&$0.05854\%$\\
\hline
  \end{tabular}
  \npnoround
  \caption{\label{tab:convergence-twobody1} Convergence of the two-body matrix
    elements for the meson-exchange kernel of \threeHe
    Compton scattering with photon helicities $(\lambda^\prime,\lambda)$ for
    potentials Idaho \NXLO{3}{}$+3$NFb and
    AV18$+$UIX in the ``density'' approach
    with $\jrel\le 1$ up to $\jrel\le 4$ at
    $\omegacm=50\;\MeV,\thetacm=30^\circ$, where mostly diagonal matrix
    elements are probed. The value for $\jrel\le4$ is in $\fm^3$ and
    normalised as in eq.~\eqref{eq:twobodynorm}.  Matrix elements 
    with $M^\prime\ne M$ are not
    shown as they have magnitudes  $<2\times10^{-7}\;\fm^3$ and are hugely
    sensitive to numerical noise but do not
    bear on any observable.  Matrix elements with $M^\prime=-\half$ 
    follow from those quoted by time-reversal invariance, see
    eq.~\eqref{eq:onebodysymmetry}. See text and caption to
    table~\ref{tab:convergence-onebody1} for further details.}
\end{table}

\begin{table}[!t]
  \small\centering
  \begin{tabular}{|c|
   >{{\nprounddigits{5}}}n{2}{5}|
    >{{\nprounddigits{2}}}n{3}{2}:
    >{{\nprounddigits{2}}}n{3}{2}:
    >{{\nprounddigits{2}}}n{3}{2}||
   >{{\nprounddigits{5}}}n{2}{5}|
    >{{\nprounddigits{2}}}n{3}{2}:
    >{{\nprounddigits{2}}}n{3}{2}:
    >{{\nprounddigits{2}}}n{3}{2}|}
\multicolumn{1}{c}{}
    &\multicolumn{8}{c}{$\omegacm=120\;\MeV,\thetacm=165^\circ$}\\
    \cline{2-9}
    \multicolumn{1}{c|}{}&
    \multicolumn{4}{c||}{Idaho \NXLO{3}{}$+3$NFb}
    &\multicolumn{4}{c|}{AV18$+$UIX}\\\hline
    \stretchboth $\,\{M^\prime,M;\lambda^\prime,\lambda\}$
    &\hspace*{\fill}{value [$\fm^3$]}\hspace*{\fill}
    &\hspace*{\fill}{$\fs1-\frac{\jrel\le1}{\jrel\le4}$}\hspace*{\fill}
    &\hspace*{\fill}{$\fs1-\frac{\jrel\le2}{\jrel\le4}$}\hspace*{\fill}
    &\hspace*{\fill}{$\fs1-\frac{\jrel\le3}{\jrel\le4}$}\hspace*{\fill}
    &\hspace*{\fill}{value [$\fm^3$]}\hspace*{\fill}
    &\hspace*{\fill}{$\fs1-\frac{\jrel\le1}{\jrel\le4}$}\hspace*{\fill}
    &\hspace*{\fill}{$\fs1-\frac{\jrel\le2}{\jrel\le4}$}\hspace*{\fill}
    &\hspace*{\fill}{$\fs1-\frac{\jrel\le3}{\jrel\le4}$}\hspace*{\fill}
    \\\hline\hline
\linel
 &$-.0014916$&$0.72406\%$&$0.13408\%$&$0.02682\%$
&$-.001898$&$1.32772\%$&$0.72708\%$&$0.06322\%$\\
 \linem
 &$-.102396$&$0.87464\%$&$0.18732\%$&$0.05216\%$
&$-.126584$&$1.33256\%$&$0.69724\%$&$0.1117\%$\\
 \linen
 &$-.102396$&$0.87464\%$&$0.18732\%$&$0.05216\%$
&$-.126584$&$1.33256\%$&$0.69724\%$&$0.1117\%$\\
 \lineo
 &$-.0014916$&$0.72406\%$&$0.13408\%$&$0.02682\%$
&$-.001898$&$1.32772\%$&$0.72708\%$&$0.06322\%$\\
\hline
  \end{tabular}
  \npnoround
  \caption{\label{tab:convergence-twobody2} Convergence of the two-body matrix
    elements as in table~\ref{tab:convergence-twobody1}, but at
    $\omegacm=120\;\MeV,\thetacm=165^\circ$, where off-diagonal matrix
    elements are probed more strongly. See also text and captions to
    tables~\ref{tab:convergence-onebody1} and~\ref{tab:convergence-twobody1}
    for further details.}
\end{table}

There is clear convergence: even the $\jrel\le1$ answers make up $>98.6\%$ of
those with $\jrel\le4$.  The progression is monotonic for all quantum numbers
and kinematic points. There are no stark differences between small and large
momentum-transfers. With $\jrel\le4$, all matrix elements are known to
$\lesssim0.1\%$, which is far better than for the one-body elements. There
also appears to be much more sensitivity to short-range details, as the
differences between answers using AV18$+$UIX or the chiral Idaho potential are
of the order of $25\%$, again at all energies, angles and quantum
numbers. This sensitivity of two-body matrix elements to the $\N\N$
potential---and in particular to the amount of $sd$-mixing---was already
observed for the deuteron~\cite{Beane:2002wn}.  Ref.~\cite{Hildebrandt:2005iw}
showed that including rescattering effects completely removes the potential
dependence as $\omega \to 0$, since it restores the Thomson limit. The
dependence on the $\N\N$ potential is then also much reduced at
$\omega \approx 100$ MeV: to only about $0.5\%$ in the cross section, with
similar reductions in other observables~\cite{Griesshammer:2013vga}.

\subsection{Comparison of the Two Approaches}
\label{sec:comparison}
 
The ``traditional'' and density approaches encode the same Physics. One might
thus expect that they lead to identical results.  However, even if both
calculations were perfectly converged individually, we would not expect
perfect agreement with previous publications. The main reasons for remaining
discrepancies are somewhat subtle:
\begin{enumerate}
\item Due to a decade and a half of advances in computing power, the wave
  functions used in the densities approach have finer momentum-spaced grids
  and correspondingly smaller interpolation errors.

\item The ``traditional'' code's wave functions were obtained from Faddeev
  calculations in momentum space~\cite{Nogga:1997mr,Epelbaum:2002vt} in a
  parametrisation that has since been superseded.

\item There are small differences in the $\N\N$ and 3$\N$ potentials between
  the code which produced the traditional wave functions $15$ years ago and
  the new implementation to construct the densities. This includes slightly
  different numerical values for the two-pion and three-nucleon interactions.
  
\item In the traditional approach, the struck nucleon in the one-body code was
  the one labelled as ``$1$'' (the ``innermost''), not ``$3$'' (the ``outermost'').  Therefore, truncation at a fixed $\jrel$
  does not mean the same thing in the two approaches.
\end{enumerate}
We do not expect any of these issues to affect the results by more than
$1$\%. In previous publications, the goal was to achieve a numerical accuracy
which was better than the thickness of the lines in plots of observables, and
considerably smaller than the overall accuracy of roughly $\lesssim3\%$ of the
\ChiEFT expansion; see detailed discussion in
ref.~\cite[sect.~2.4.3]{Margaryan:2018opu}. Physics, rather than numerics, was
the focus, and including \threeHe channels up to $\jrel = 2$ achieved the
goal. Now, we compare instead our new approach to Compton scattering with
those previous $\jrel^{\rm max}=2$ results. The points of difference listed
above could, in principle, be improved in the traditional code. But we have
decided to ``retire'' that inefficient implementation after verifying that the
densities approach reproduces it to acceptable accuracy. Thus, our standard
for agreement between the two approaches is 1\%.

Finally, we note that comparing the one-body insertions $\sigma_\mu^{(\N)}$
with $\jrel=2,3$ in the two approaches helped us diagnose a mistake in the
implementation of the ``traditional approach'' which is discussed in
app.~\ref{sec:previous}. There was a noticeable disagreement between the two
approaches, until that was corrected.

\subsubsection{Comparison of One-Body Matrix Elements}
\label{sec:comparison1b}

\begin{table}[!t]
  \small\centering
  \begin{tabular}{|l|@{\hspace{0.2ex}}c||
    >{{\nprounddigits{4}}}n{2}{4}|>{{\nprounddigits{2}}}n{3}{2}||
    >{{\nprounddigits{4}}}n{2}{4}|>{{\nprounddigits{2}}}n{3}{2}?{2pt}
    >{{\nprounddigits{4}}}n{2}{4}|>{{\nprounddigits{2}}}n{3}{2}||
    >{{\nprounddigits{4}}}n{2}{4}|>{{\nprounddigits{2}}}n{3}{2}|}
    \multicolumn{2}{c}{}
    &\multicolumn{8}{c}{$\omegacm=50\;\MeV,\thetacm=30^\circ$}\\
    \hline
    \multicolumn{2}{|c||}{\multirow{2}{*}{insertion}}&
    \multicolumn{4}{c?{2pt}}{Idaho \NXLO{3}{}$+3$NFb}&
    \multicolumn{4}{c|}{AV18$+$UIX}\\\cline{3-10}
    \multicolumn{2}{|c||}{}&
    \multicolumn{2}{c||}{proton}&\multicolumn{2}{c?{2pt}}{neutron}&
    \multicolumn{2}{c||}{proton}&\multicolumn{2}{c|}{neutron}\\\hline
    \stretchboth$\;\sigma_\mu\hqm$&$\,\{M^\prime,M\}$
&\hspace*{\fill}{value}\hspace*{\fill}&\hspace*{\fill}{rel.dev.}\hspace*{\fill}
&\hspace*{\fill}{value}\hspace*{\fill}&\hspace*{\fill}{rel.dev.}\hspace*{\fill}
&\hspace*{\fill}{value}\hspace*{\fill}&\hspace*{\fill}{rel.dev.}\hspace*{\fill}
&\hspace*{\fill}{value}\hspace*{\fill}&\hspace*{\fill}{rel.dev.}\hspace*{\fill}
    \\\hline\hline
\linea
 &$1.96775$&$-0.119886\%$&$.98607$&$-0.0270434\%$ 
&$1.95596$&$0.149338\%$&$.979116$&$0.0823541\%$\\
 \hline 
 \lined
 &$-.0398394$&$0.0650536\%$&$.880923$&$0.544949\%$ 
&$-.0496441$&$3.7038\%$&$.86103$&$0.7563\%$\\
 \hline
 \linef
 &$-.0375884$&$0.22641\%$&$.88101$&$0.545729\%$ 
&$-.0476213$&$3.90735\%$&$.861105$&$0.756753\%$\\
 \hline\lineg
 &$-.0377524$&$0.220308\%$&$.881001$&$0.545365\%$ 
&$-.0477713$&$3.90189\%$&$.861097$&$0.756382\%$\\
 \lineh
 &$.000603138$&$-2.62971\%$&$.000023516$&$8.36786\%$ 
&$.000542015$&$-1.08856\%$&$.000019992$&$5.98584\%$\\
\hline
  \end{tabular}
  \npnoround
  \caption{\label{tab:comparison-onebody1} Comparison of the independent
    nonzero
    one-body matrix elements in \threeHe with insertions $3\sigma_\mu^{(\N)}$
    for the proton and neutron, for potentials Idaho \NXLO{3}{}$+3$NFb and
    AV18$+$UIX.  The column ``rel.~dev.'' denotes the relative
    difference, as defined in eq.~\eqref{eq:relativedifference}, between the 
    ``density'' and ``traditional''
    approach with $\jrel\le2$ at
    $\omegacm=50\;\MeV,\thetacm=30^\circ$, where mostly diagonal matrix
    elements are probed. See text and caption to
    table~\ref{tab:convergence-onebody1} for further details.}
\end{table}

\begin{table}[!ht]
  \small\centering
  \begin{tabular}{|l|@{\hspace{0.2ex}}c||
   >{{\nprounddigits{4}}}n{2}{4}|>{{\nprounddigits{2}}}n{3}{2}||
   >{{\nprounddigits{4}}}n{2}{4}|>{{\nprounddigits{2}}}n{3}{2}?{2pt}
   >{{\nprounddigits{4}}}n{2}{4}|>{{\nprounddigits{2}}}n{3}{2}||
   >{{\nprounddigits{4}}}n{2}{4}|>{{\nprounddigits{2}}}n{3}{2}|}
    \multicolumn{2}{c}{}
    &\multicolumn{8}{c}{$\omegacm=120\;\MeV,\thetacm=165^\circ$}\\
    \hline
    \multicolumn{2}{|c||}{\multirow{2}{*}{insertion}}&
    \multicolumn{4}{c?{2pt}}{Idaho \NXLO{3}{}$+3$NFb}&
    \multicolumn{4}{c|}{AV18$+$UIX}\\\cline{3-10}
    \multicolumn{2}{|c||}{}&
    \multicolumn{2}{c||}{proton}&\multicolumn{2}{c?{2pt}}{neutron}&
    \multicolumn{2}{c||}{proton}&\multicolumn{2}{c|}{neutron}\\\hline
    \stretchboth$\;\sigma_\mu\hqm$&$\,\{M^\prime,M\}$
&\hspace*{\fill}{value}\hspace*{\fill}&\hspace*{\fill}{rel.dev.}\hspace*{\fill}
&\hspace*{\fill}{value}\hspace*{\fill}&\hspace*{\fill}{rel.dev.}\hspace*{\fill}
&\hspace*{\fill}{value}\hspace*{\fill}&\hspace*{\fill}{rel.dev.}\hspace*{\fill}
&\hspace*{\fill}{value}\hspace*{\fill}&\hspace*{\fill}{rel.dev.}\hspace*{\fill}
    \\\hline\hline
\linea
 &$.991314$&$-0.064905\%$&$.542401$&$0.759762\%$ 
&$1.02295$&$-0.431101\%$&$.562204$&$1.13293\%$\\
 \hline 
 \lined
 &$.0385298$&$5.1566\%$&$.479331$&$0.981337\%$ 
&$.0251495$&$2.58459\%$&$.482207$&$1.3259\%$\\
 \hline
 \linef
 &$.0404145$&$4.91684\%$&$.479425$&$0.981484\%$ 
&$.0268549$&$2.45946\%$&$.482287$&$1.32643\%$\\
 \hline\lineg
 &$-.0683227$&$-2.8859\%$&$.47401$&$0.973121\%$ 
&$-.0715409$&$-0.0798323\%$&$.477675$&$1.29542\%$\\
 \lineh
 &$.0143156$&$0.0149515\%$&$.000713088$&$1.73223\%$ 
&$.0129541$&$0.613849\%$&$.000607177$&$4.53694\%$\\
\hline
  \end{tabular}
  \npnoround
  \caption{\label{tab:comparison-onebody2} Comparison of the  one-body matrix
    elements as  in table~\ref{tab:comparison-onebody1}, but at
    $\omegacm=120\;\MeV,\thetacm=165^\circ$, where off-diagonal matrix
    elements are probed more strongly.}
\end{table}

In ref.~\cite{Margaryan:2018opu}, we studied numerical convergence only of the
overall Compton one-body matrix elements. These are dominated by $\gamma\N$
interactions which do not involve nucleon-structure effects such as the
nucleon polarisabilities, \eg~the Thomson term (insertion of
$\sigma_0^{(\N)}\equiv\mathbbm{1}$) as well as the minimal electric and
magnetic-moment couplings to the nucleon. Therefore, including all partial
waves with $\jrel \le2$ in the ``traditional'' one-body matrix elements
sufficed for convergence of Compton matrix elements to within $0.5\%$ at the
highest energy and momentum-transfer we considered. The cross section was then
numerically converged at about $1.2\%$ or $0.35$~nb/sr there, and considerably
better elsewhere.

In contradistinction, we now compare the \threeHe matrix elements at a given
energy and angle for each of the $8$ insertions $\sigma_\mu^{(\N)}$, \ie~we
look at more than just overall one-body Compton matrix elements. We reiterate
that in what follows, we do not compare to the ``fully converged'' density
results with $\jrel\le5$ of sect.~\ref{sec:convergence1b}, but to those which
use the same $\jrel\le2$ as the ``traditional'' approach.

Tables~\ref{tab:comparison-onebody1} and~\ref{tab:comparison-onebody2} show
that the two methods agree very well. The matrix elements of order $1$ agree
to $\approx1\%$ and better. Those of order $10^{-[1\dots2]}$ show somewhat
more variance, with a relative deviation as defined in
eq.~\eqref{eq:relativedifference} of $\lesssim5\%$ at the largest momentum
transfers. That is still less than $0.5\%$ of the spin-helicity contributions
with the largest magnitudes at a given energy and angle; see criterion in
eq.~\eqref{eq:differencerelativetomax}. Matrix elements with magnitudes of
order $10^{-4}$ and smaller show up to $8\%$ relative variation, or
$\lesssim0.01\%$ relative to the matrix element with the largest magnitude.
The agreement is usually better for the chiral potential than for AV18$+$UIX,
as its harder core needs a finer interpolation in densities and wave
functions.

\subsubsection{Comparison of Two-Body Matrix Elements}
\label{sec:comparison2b}

In the ``traditional'' two-body matrix elements, a maximum total angular
momentum of the $(12)$ subsystem $\jrel\le1$ provides a reasonable compromise
between runtime and numerical accuracy. With an increase to $\jrel\le2$, the
matrix elements change by barely more than $0.7\%$ for either choice of
potential---even at the highest energies and momentum transfers considered.
However, the runtime increases nearly tenfold, from CPU-hours to days per
energy and angle, and even then their numerical accuracy is not quite as good
as that of the density method; see the enumeration at the beginning of
sect.~\ref{sec:comparison}. Certainly then, going to $\jrel=3$ is not
worthwhile for the two-body matrix elements we consider here. Therefore, we
decided to compare results for $\jrel\le2$.

Only matrix elements which are negligible (namely $<10^{-6}$ relative to the
biggest ones) show substantial relative differences upon the inclusion of
channels with $\jrel=2$.  All this is consistent with the pattern which
emerged in the convergence-check of the density results as well; see
sect.~\ref{sec:convergence2b}.

\begin{table}[!t]
  \small\centering
  \setlength{\tabcolsep}{1.7pt}
  \renewcommand{\arraystretch}{1.2}
  \providecommand{\stretchboth}{\rule[-1.5ex]{0pt}{4.5ex}}
  \providecommand{\stretchup}{\rule[0ex]{0pt}{3ex}}
  \providecommand{\stretchdown}{\rule[-1.5ex]{0pt}{0ex}}
  \newcolumntype{?}[1]{!{\vrule width #1}}
  \npdecimalsign{.}
  \nprounddigits{5} 
  \npthousandthpartsep{} \begin{tabular}{|@{\hspace{0.2ex}}c|
   >{{\nprounddigits{5}}}n{2}{5}|
    >{{\nprounddigits{1}}}n{3}{1}||
   >{{\nprounddigits{5}}}n{2}{5}|
    >{{\nprounddigits{1}}}n{3}{1}?{2pt}
   >{{\nprounddigits{5}}}n{2}{5}|
    >{{\nprounddigits{1}}}n{3}{1}||
   >{{\nprounddigits{5}}}n{2}{5}|
    >{{\nprounddigits{1}}}n{3}{1}|}
   \multicolumn{1}{c}{}
    &\multicolumn{4}{c?{2pt}}{$\omegacm=50\;\MeV,\thetacm=30^\circ$}
    &\multicolumn{4}{c}{$\omegacm=120\;\MeV,\thetacm=165^\circ$}\\
    \cline{2-9}\multicolumn{1}{c}{}
    &\multicolumn{2}{|c||}{Idaho \NXLO{3}{}$+3$NFb}
    &\multicolumn{2}{c?{2pt}}{AV18$+$UIX}
    &\multicolumn{2}{c||}{Idaho \NXLO{3}{}$+3$NFb}
    &\multicolumn{2}{c|}{AV18$+$UIX}\\
                                             \hline
    \stretchboth$\{M^\prime,M;\lambda^\prime,\lambda\}$
&\hspace*{\fill}{value [$\fm^3$]}\hspace*{\fill}&\hspace*{\fill}{rel.dev.}\hspace*{\fill}
&\hspace*{\fill}{value [$\fm^3$]}\hspace*{\fill}&\hspace*{\fill}{rel.dev.}\hspace*{\fill}
&\hspace*{\fill}{value [$\fm^3$]}\hspace*{\fill}&\hspace*{\fill}{rel.dev.}\hspace*{\fill}
&\hspace*{\fill}{value [$\fm^3 $]}\hspace*{\fill}&\hspace*{\fill}{rel.dev.}\hspace*{\fill}
    \\\hline\hline
\linel
 &$-.0713236$&$0.14946\%$
&$-.0934346$&$0.18558\%$ 
 &$-.0014896$&$0.\%$
&$-.0018842$&$0.24414\%$\\
 \linem
 &$-.0054276$&$0.30216\%$
&$-.007024$&$0.29612\%$ 
 &$-.102204$&$0.84752\%$
&$-.125701$&$0.7844\%$\\
 \linen
 &$-.0054276$&$0.29848\%$
&$-.0070238$&$0.29044\%$ 
 &$-.102204$&$0.84752\%$
&$-.125701$&$0.7844\%$\\
 \lineo
 &$-.0713236$&$0.14946\%$
&$-.0934346$&$0.18558\%$ 
 &$-.0014896$&$0.\%$
&$-.0018842$&$0.24414\%$\\
\hline
  \end{tabular}
  \npnoround
  \caption{\label{tab:comparison-twobody1} Comparison of two-body matrix
    elements in the 
    ``density'' approach and the ``traditional''
    approach for potentials Idaho \NXLO{3}{}$+3$NFb and
    AV18$+$UIX with $\jrel\le2$ at
    $\omegacm=50\;\MeV,\thetacm=30^\circ$ (where mostly diagonal matrix
    elements are probed) and $\omegacm=120\;\MeV,\thetacm=165^\circ$ (where
    off-diagonal matrix elements are probed more strongly). See also text and captions
    to tables~\ref{tab:comparison-onebody1} and~\ref{tab:convergence-twobody1}
    for further details.} 
\end{table}
Bearing in mind that the numerical treatment of the $(12)$ subsystem is
identical and indeed uses the same code, it is not surprising that the
CPU-time for two-body matrix elements increases roughly by a factor of $10$
from $\jrel\le1$ to $\le2$ in both the ``traditional'' and density
approach. But the ``traditional'' matrix-element evaluation is about 20 times
slower in each case. It spends the vast majority of its time on the part which
is encoded in the two-body densities that serve as input in the ``densities''
approach. A minor price to pay is that the two-body densities are very big;
see the end of sect.~\ref{sec:twobodydensity}.

As can be seen in table~\ref{tab:comparison-twobody1}, ``traditional'' and
density approach agree to better than $0.3\%$ for the chiral potential, and
better than $1\%$ for AV18$+$UIX even at the higher energies and momentum
transfers, based on the criterion of eq.~\eqref{eq:relativedifference}. That
is very close to the difference between the results at $\jrel\le2$ and the
converged result; \cf~tables~\ref{tab:convergence-twobody1}
and~\ref{tab:convergence-twobody2}.

We therefore conclude that the one- and two-body matrix elements agree very
well in the two approaches---namely to within the 1\% expected after the
discussion of difference between the two implementations in the opening of
sect.~\ref{sec:comparison}.

\section{Summary and Outlook}
\label{sec:conclusion}

We introduced a transition-density method that employs pre-computed one- and
two-body densities in the evaluation of elastic processes in which momentum is
transferred to an $A$-nucleon system, and used Compton scattering on \threeHe
as a test case.  Extensions to charge-transfer or inelastic processes, to
incorporate few-body transition densities with $n\ge3$ nucleons active in the
reaction process and to other targets are conceptually relatively
straightforward.

The method has several attractive features. Producing transition densities is
the computationally most demanding aspect of the method---but once produced,
they can be applied to a host of reactions. Therefore, their quality for a
particular nucleus can be extensively benchmarked against known processes, and
computational resources and development can be focused on densities. On the
other hand, a particular reaction kernel involves only those nucleons which
interact with the probe, and not the spectators. Therefore, the quality of a
kernel can be benchmarked across different nuclei.  Once the pertinent one-
and few-body densities have been calculated for a new nucleus, only small
changes in existing matrix-element calculations are required, \eg~because of
the different quantum numbers of the particular nuclear ground states.  The
computational effort needed to go from a given kernel and a given set of
densities to interaction matrix elements is therefore hardly different for an
arbitrary nucleus than it is for, say, \threeHe.

In our example of coherent Compton scattering, the single- and two-nucleon
Compton kernel are already available in the one- and two-nucleon Hilbert
spaces, respectively. The improvement achieved here therefore opens the way to
using the same operators for Compton matrix elements on other nuclei with
$A\ge3$, like \fourHe. Once we have densities for heavier targets, Compton
matrix elements can be produced quickly.

In Compton scattering on \threeHe, the densities-based method also turned out
to be markedly faster than the calculational strategy employed
previously. This allowed for detailed convergence studies with an
unprecedented number of partial waves. These show very good numerical
convergence even for a comparatively hard underlying potential like AV$18$
with the Urbana-IX $3\N$ interaction. Such investigations would have been
prohibitively expensive in the previously-used method. The new method produces
results which agree with the traditional ones at the expected level for both a
``hard'' and ``soft'' $\N\N$ and $3\N$ interaction, taking into account the
limitations of the calculational aspects of the old implementation. Having
such similar results from two largely different codes and methodological
approaches makes us confident that the coding and numerics is fully
understood.

It was not our goal here to provide predictions that can be compared with
data. Instead, we wanted to validate our new, computationally less intensive,
densities-based method.  Nevertheless, it is useful to provide some
context---and perhaps inject some caution---regarding the numbers presented
above. We thus recapitulate part of the discussion of
ref.~\cite{Margaryan:2018opu} here. For our high-energy results at
$\omega\approx120\;\MeV$, the kernel and \threeHe state are complete up to and
including chiral order $e^2\delta^3$, where $\delta\approx0.4$ is the typical
size of the expansion parameter~\cite{Beane:1999uq,Pascalutsa:2002pi}. The
first omitted terms can thus be estimated as $\calO(e^2\delta^4)\lesssim3\%$
in the amplitudes, or twice that in a cross section. These omitted terms
include ``rescattering'' as discussed in the Introduction and
sect.~\ref{sec:overview}: the coherent propagation of all target nucleons
between photon absorption and emission. As $\omega\to0$, however, rescattering
becomes dominant and the chiral power counting changes~\cite{Chen:1998ie,
  Griesshammer:2012we}.

As the photon energy decreases, the importance of rescattering increases
gradually. In the deuteron, rescattering accounts indeed for about $3\%$ of
the cross section at $\omega\approx90\;\MeV$, in line with the analysis of the
previous paragraph, but it is a $20\%$ effect at $\omega\approx45\;\MeV$ (see
fig.~5.3 of ref.~\cite{Griesshammer:2012we}).  This is consistent with its
effect scaling as $1/(\omega-\omega_\mathrm{coll})$, where
$\omega_\mathrm{coll}$ is the scale at which resummation becomes
unavoidable. We suspect $\omega_\mathrm{coll} \sim \mpi^2/\MN$ with $\MN$ the
nucleon mass, since $\mpi$ is the typical momentum scale of the
interaction~\cite{Beane:1999uq,Hildebrandt:2005iw}. This scale would then not
be especially nucleus dependent---albeit we expect it is somewhat higher in
\threeHe than in the deuteron. This suggests that the results presented above
have a theoretical uncertainty of about $20\%$ at
$\omega\approx50\;\MeV$. While this is clearly insufficient for high-accuracy
comparisons with data, even a $20\%$ uncertainty can suffice to encourage and
guide experimental planning and data taking. A more reliable quantification of
the uncertainty from rescattering needs a computation of coherent-$A$-body
rescattering in \threeHe. This is under way~\cite{Johannes}.
  
Another obvious next step is the calculation of densities for \fourHe and of
its elastic Compton cross section at energies up to about $120\;\MeV$. Those
can then be compared with the recent data from
HI$\gamma$S~\cite{Sikora:2017rfk, Li:2019irp} with the goal of extracting
high-accuracy values for the nucleon polarisabilities. As the two-nucleon
operators considered here are quite similar to those for dark-matter
scattering on nuclei, we also intend to use previous dark-matter-\fourHe
scattering calculations~\cite{Korber:2017ery,Andreoli:2018etf} as benchmarks
for an evaluation with pre-computed densities.

Targets beyond \fourHe are, again, not computationally more costly, once
densities have been computed. Compton scattering off heavier targets, like
${}^6$Li~\cite{Myers:2014qaa}, will presumably require densities from No-Core
Shell Model wave functions~\cite{Maris:2016wrd}. In this regard, the approach
adopted here has much in common with the recent work of Burrows \etal, where
single-body densities were used to compute nucleon-nucleus optical potentials
for \fourHe, ${}^6$He, ${}^{12}$C and
${}^{16}$O~\cite{Burrows:2017wqn,Burrows:2018ggt, Burrows:2020qvu}.  In that
case, however, a different, non-local, density enters, since the density is
folded with the nucleon-nucleon $T$-matrix, and not with an operator that is
local in coordinate space. Two-body densities were not considered in
ref.~\cite{Burrows:2017wqn}, either; they would presumably be required in a
calculation of $3\N$-interaction corrections to the optical potential.

So far, we produced only densities for one-body operators which depend at most
linearly on the total cm momentum of the nucleus, besides the dependence on
the momentum-transfer. For now, the two-body densities require two-body
operators that are independent of the total momentum of the nucleus.  The
extension to higher-rank dependence on the momentum of the nucleus is
straightforward if needed.

The transition-density method has applications well beyond Compton scattering;
a cornucopia of processes can be computed with the densities introduced
here. Any elastic scattering process in which a probe interacts perturbatively
with \threeHe can be evaluated using our densities, provided the pertinent
reaction kernels are written as momentum-space interactions with only one or
two active nucleons, and as long as they fulfil the criteria stated in the
previous paragraph. For example, the one-body densities are exactly those
needed to compute single-nucleon operator contributions to electron scattering
on a nucleus. Likewise, the two-body
densities are sufficiently general that they can be used to compute
exchange-current corrections to the form factors from two-body operators in
momentum space.

Practitioners interested in convoluting momentum-space operators with our
\threeHe densities can find them at
\url{https://datapub.fz-juelich.de/anogga}. Densities are provided for AV$18$
with Urbana-IX $3\N$~\cite{Wiringa:1994wb, Pudliner:1995wk} interaction and
the chiral Idaho \NXLO{3} potential at cutoff $500\;\MeV$~\cite{Entem:2003ft}
with the ``$\mathcal{O}(Q^3)$'' \ChiEFT $3\N$ interaction of variant ``b'' of
ref.~\cite{Nogga:2005hp}. We will provide densities based on other modern,
sophisticated potentials in the future and encourage practitioners to contact
us with requests for further extensions.


\section*{Acknowledgements}

We thank Mike Birse for useful input at a couple of stages of this work.  DRP
thanks Charlotte Elster and Matt Burrows for informative discussions. The
stimulating environment and financial support of the INT in Seattle came at a
critical juncture of this research. We are therefore grateful to the
organisers and participants of the INT ``Programme 18-2a: Fundamental Physics
with Electroweak Probes of Light Nuclei'' and INT workshop ``From Nucleons to
Nuclei: Enabling Discovery for Neutrinos, Dark Matter And More''.
HWG acknowledges the warm hospitality and financial support of Ohio
University, the University of Manchester and Forschungszentrum J\"ulich which
was instrumental for this research. Likewise, AN is grateful for the warm
hospitality and financial support of Ohio University. DRP is grateful for the
warm hospitality of the IKP Theoriezentrum, Darmstadt.
This work was supported in part by the US Department of Energy under contract
DE-SC0015393 (HWG) and DE-FG02-93ER-40756 (DRP), by the UK Science and
Technology Facilities Council grant ST/P004423/1 (JMcG), by the ExtreMe Matter
Institute EMMI at the GSI Helmholtzzentrum f\"ur Schwerionenphysik, Darmstadt,
Germany (DRP), and by the Deutsche Forschungsgemeinschaft and the Chinese
National Natural Science Foundation through funds provided to the Sino-German
CRC 110 ``Symmetries and the Emergence of Structure in QCD'' (AN; DFG grant
no. TRR~110; NSFC grant no. 11621131001). Additional funds for HWG were
provided by an award of the High Intensity Gamma-Ray Source \HIGS of the
Triangle Universities Nuclear Laboratory TUNL in concert with the Department
of Physics of Duke University, and by George Washington University: by the
Office of the Vice President for Research and the Dean of the Columbian
College of Arts and Sciences; by an Enhanced Faculty Travel Award of the
Columbian College of Arts and Sciences. His research was conducted in part in
GW's Campus in the Closet.
The computations of nuclear densities were performed on \textsc{Jureca} and the
\textsc{Jureca-Booster} of the J\"ulich Supercomputing Centre (J\"ulich, Germany).
\newpage

\appendix

\section{Comment on Prior Compton Calculations on \threeHe}
\label{sec:previous}

Our previous strategy for the computation of \threeHe matrix elements of the
Compton operators was based on the photodissociation calculation of
ref.~\cite{Kotlyar:1999an}. The analogous integrals for one-body and two-body
operator contributions to the matrix elements were performed without splitting
them into reaction-mechanism and density parts. While they were factorised
into a piece involving the nucleons taking part in the reaction and the matrix
element of the spectator $\delta$-distribution, the efficiency of defining
densities that refer only to the quantum numbers of the active nucleons was
not noticed; see refs.~\cite{Choudhury:2007bh, Shukla:2018rzp,
  Shukla:2008zc,ShuklaPhD, Margaryan:2018opu} for details.

In the course of this study, we found that refs.~\cite{Choudhury:2007bh,
  Shukla:2018rzp, Shukla:2008zc,ShuklaPhD, Margaryan:2018opu} contain a flaw
in the reasoning leading to the original equations corresponding to
eq.~\eqref{eq:targetme}, which in turn led to incorrect numerical
implementations of the one-body part. The struck nucleon in the one-body part
was considered to be not nucleon $3$ but one of the nucleons of the $(12)$
sub-system. Therefore, rather than ${\hat O}_{\lambda^\prime \lambda}^{1B}(3)$
as in sect.~\ref{sec:onebodydensity}, the operator
${\hat O}_{\lambda^\prime \lambda}^{1B}(1)$ was considered and replaced by
$\frac{1}{2} [{\hat O}_{\lambda^\prime \lambda}^{1B}(1) + {\hat
  O}_{\lambda^\prime \lambda}^{1B}(2)]$
in the course of defining operators on the space of two-nucleon states.
However, this replacement cannot be done at the level of the spin and isospin
operators because the momentum assignment for the post-collision state differs
depending on whether the struck particle is nucleon $1$ or nucleon $2$.
  
That error in refs.~\cite{Choudhury:2007bh, Shukla:2018rzp,
  Shukla:2008zc,ShuklaPhD} was also present in our recent evaluation of
$\gamma$\threeHe scattering~\cite{Margaryan:2018opu}. It means that those
works missed contributions at nonzero momentum-transfer, where the
Compton-scattering collision induced a transition that changed either the spin
or the isospin of the $\N\N$ state, but not both.
  
Fortunately, the numerical effect on observables is very small.  For the
neutron, this changes the matrix elements with insertions of $3\sigma_\mu$ by
$\le3\%$, except for about $16\%$ in $\sigma_x$ at the highest energy and
momentum transfer we consider $(\omegacm=120\;\MeV,\thetacm=165^\circ)$. The
change is more pronounced for the proton, where it can amount to a factor of
about $4.5$ in $\sigma_x$ at that point and exceeds $10\%$ even at small
$(\omegacm,\thetacm)$. This might seem to imply big changes of the one-body
amplitudes for the spin-polarisabilities.  But the proton spins inside
\threeHe are mostly paired to spin-zero, so there is hardly any sensitivity to
the mistake in matrix elements of the proton spin. The effect is also shielded
for the neutron spin. Even at the ``high'' energy and momentum-transfer tested
here, the effect of the neutron's spin-polarisabilities is $\approx 10$\% in
the amplitude. Matrix elements of the neutron's spin do not play a big role,
and even a 16\% error in them would only be a 1.6\% error in matrix elements.
This is associated with the fact that the biggest contributions to \threeHe
Compton scattering for $50\;\MeV\lesssim\omegacm\lesssim120\;\MeV$ come from
interactions with the two charged protons. These do not change the nucleon
spin and are hence proportional to insertions of
$\mathbbm{1}\equiv\sigma_0$---and such matrix elements are changed by less
than $0.2\%$ of the largest magnitude of all one-body matrix elements, see
eq.~\eqref{eq:differencerelativetomax}. Therefore, we were able to find the
error only when we zoomed in on a detailed comparison between the
``traditional'' and ``density'' approach; see sect.~\ref{sec:comparison}. In
the ``traditional" results quoted in the body of the paper this error is, of
course, corrected.

In almost all cases, this mistake for the matrix elements with insertions
$\sigma_\mu^{(\N)}$ only minimally alters the plots of both magnitudes and
sensitivities of observables in ref.~\cite{Margaryan:2018opu}. The cross
section as well as the double-asymmetries
$T_{11}^\mathrm{circ}\equiv\Sigma_{2x}$ and
$T_{10}^\mathrm{circ}\equiv\Sigma_{2z}$ change by $<1\%$ at $50\;\MeV$, and by
$\lesssim3\%$ at $120\;\MeV$, where asymmetries exceed $0.1$. For
$\Sigma^\mathrm{lin}\equiv\Sigma_{30}$, the thickness of the line is never
exceeded ($<1\%$). To put this into perspective, the variation from using
different \threeHe wave functions is at all energies and angles at least a
factor five bigger than the change from this error. In
ref.~\cite{Margaryan:2018opu}, wave-function dependence was, in turn,
estimated to be substantially smaller than the sum of all residual theoretical
uncertainties. Therefore, we refrain from amending or updating the
presentations of refs.~\cite{Choudhury:2007bh, Shukla:2018rzp, Shukla:2008zc,
  Margaryan:2018opu}. Their conclusions are unchanged.

\section{Symmetries of Matrix Elements}
\label{app:symmetries}

We now derive the symmetries that relate different matrix elements of the one-
and two-nucleon operators in sects.~\ref{sec:comparison}
and~\ref{sec:convergence} by considering an insertion
$\sigma_\mu\, \phasefactor$, which is independent of $\kv$ (\ie~$K=\kappa=0$).
Note that we have employed the Jacobi-coordinate-space representation of the
momentum-conservation relation~\eqref{eq:delta-in-3} here to define the
operator insertion. Hence, the plane wave deposits momentum into the Jacobi
co-ordinate of nucleon 3. Also, $\sigma_\mu^{(\N)}$ acts not on the \threeHe
nucleus as a whole but only on the single ``active'' nucleon $3$ at coordinate
$\vec{r}_3$.  Therefore, this set of operators cannot be represented by the
standard Pauli matrices.  However, their matrix elements do retain certain
properties of a spin-$\half$ representation.

We first prove relation \eqref{eq:symmetry-onebody}. Under time reversal,
$\mathcal{T} \sigma_i^{(\N)} \mathcal{T}^{-1}=-\sigma_i^{(\N)}$ is odd, while
$\mathcal{T} \sigma_0^{(\N)} \mathcal{T}^{-1}=\sigma_0^{(\N)}$ is even. Since
we wish to consider only operators with real matrix elements, we also note
that $\mathcal{T} \ii \sigma_y^{(\N)} \mathcal{T}^{-1}=\ii \sigma_y^{(\N)}$ is
even.

Now denoting the state $\mathcal{T}|\psi\rangle$ by $|\tilde\psi\rangle$, time
reversal invariance gives for the matrix element of some operator
$\mathcal{Q}$, see \eg~ref.~\cite{Sakurai}:
\begin{equation}
\bra  \phi | \mathcal{Q} |\psi\ket=\bra \tilde \psi | \mathcal{T} \mathcal{Q}^\dagger \mathcal{T}^{-1} |\tilde \phi\ket\;\;.
\end{equation}
If the matrix element is real, then $\bra \phi | \mathcal{Q} |\psi\ket = \bra \psi | \mathcal{Q}^\dagger |\phi\ket $, so 
\begin{equation}
\bra  \psi | \mathcal{Q}^\dagger  |\phi\ket=\bra \tilde \psi | \mathcal{T} \mathcal{Q}^\dagger \mathcal{T}^{-1} |\tilde \phi\ket\;\;.
\end{equation}
This, of course, remains true if we replace $ \mathcal{Q}^\dagger$ with
$ \mathcal{Q}$, which now serves as our starting point.  Recalling
eq.~\eqref{eq:Tsym}
\begin{equation}
   \mathcal{T}   |j,m_j\ket=\sign^{j+m_j} \;|j,-m_j \ket\;\;,
\end{equation}
we obtain
\begin{align}
  \label{eq:sigmaxz} 
  \bra \wfbra M'|\sigma_{x,z}^{(\N)}|\wf M \ket&=\sign^{2J+M'+M-1} \;\bra \wfbra
  -M'|\sigma_{x,z}^{(\N)}|\wf -M \ket\\
  \bra \wfbra M'|\sigma_0^{(\N)}|\wf M \ket&=\sign^{2J+M'+M}\; \bra \wfbra
  -M'|\sigma_0^{(\N)}|\wf -M \ket  \label{eq:sigma0}  \\
  \label{eq:isigmay}
  \bra \wfbra M'|\ii \sigma_{y}^{(\N)}|\wf M \ket&=\sign^{2J+M'+M} \;\bra \wfbra
  -M'|\ii \sigma_{y}^{(\N)}|\wf -M \ket
  \;\;. 
\end{align}
As $2(J+M)$ is even for both half-integer and integer quantum numbers, this
proves eq.~\eqref{eq:symmetry-onebody}.  Replacing
$\sigma_\mu^{(\N)}\to\sigma_\mu^{(\N)}\, \phasefactor$ does not alter this
argument since the plane wave is time-reversal even (parity guarantees that
only its real parts can contribute to the final result).  Accounting for the
multipolarity of $\kv$ ($K\ne0$), the relation is modified to:
\begin{equation}
  \label{eq:symmetry-onebody-general}
  A^{-M^\prime}_{-M}(\sigma_\mu^{(\N)};K,-\kappa,\qv)=
  \sign^{M^\prime-M+\mu+\kappa}\,A^{M^\prime}_{M}(\sigma_\mu^{(\N)};K,\kappa,\qv)\;\;. 
\end{equation}

This proves there are at most eight independent matrix elements of the
$\sigma_\mu^{(\N)}\, e^{\ii \frac{2}{3} \qv \cdot \vec{r}_3}$ operator in the
spin-$\half$ basis $(M^\prime M)$. We now discuss symmetries that eliminate
three more. First, since $\ii\sigma_y^{(\N)}$ is both real and anti-Hermitean,
both its diagonal elements must be zero:
\begin{equation}
  \label{eq:sigmayzero}
  A^{M}_M(\ii\sigma_y^{(\N)};\qv)\equiv \bra \wfbra M|\ii \sigma_y^{(\N)}\,\phasefactor|\wf M \ket=0\;\;.
\end{equation}
Furthermore $\sigma_0^{(\N)}$ is symmetric, but  its off-diagonal elements are equal and opposite by \eqref{eq:sigma0}. Hence they must be zero:
\begin{equation}
  \label{eq:sigma0zero}
  A^{-M}_M(\sigma_0^{(\N)};\qv)\equiv\bra \wfbra -M|\sigma_0^{(\N)}\,\phasefactor|\wf M \ket=0\;\;.
\end{equation}
These relations can also be established from the Lie algebra of the Pauli
operators in a two-dimensional representation that is consistent with time
reversal and in which $\sigma_x^{(\N)}$ and $\sigma_z^{(\N)}$ are real. They
do not hold for densities which explicitly depend on $\kv$ (\ie~$K\ne0$).

If the spin of the nucleon were always perfectly aligned with the spin of a
$J=\half$ nucleus, then off-diagonal matrix elements of $\sigma_z^{(\N)}$ and
diagonal matrix elements of $\sigma_x^{(\N)}$ would also be zero. These matrix
elements pick out densities for ``wrong-spin'' to ``right-spin'' transitions
(or vice versa). Specifically:
\begin{equation}
  \begin{split}
    \label{eq:explicitsigmas}
    \bra \wfbra -\tfrac{1}{2}|\sigma_z^{(\N)}\,\phasefactor
    |\wf \tfrac{1}{2} \ket&=\rho^{00; m_3^t M_T,-\frac{1}{2}, +\frac{1}{2}}_{+\frac{1}{2}, +\frac{1}{2}} - \rho^{00; m_3^t M_T,-\frac{1}{2}, +\frac{1}{2}}_{-\frac{1}{2}, -\frac{1}{2}} \;\;,\\
    \bra \wfbra \tfrac{1}{2}|\sigma_x^{(\N)}\,\phasefactor
    |\wf \tfrac{1}{2}\ket&=\rho^{00; m_3^t M_T,+\frac{1}{2}, +\frac{1}{2}}_{-\frac{1}{2}, +\frac{1}{2}}+ \rho^{00; m_3^t M_T,+\frac{1}{2}, +\frac{1}{2}}_{+\frac{1}{2}, -\frac{1}{2}} \;\; .
  \end{split}
\end{equation} 
Time-reversal alone is not enough to guarantee the equality of these two
matrix elements. But the flipping symmetry of eq.~\eqref{eq:onebodyflipping}
means that the first terms of each line of eq.~\eqref{eq:explicitsigmas} are
equal. Using flipping symmetry in conjunction with time-reversal
\eqref{eq:onebodysymmetry} shows that the second terms are equal,
too. Therefore, the off-diagonal matrix elements of $\sigma_z^{(\N)}$ and
diagonal ones of $\sigma_x^{(\N)}$ are identical:
\begin{equation}
  \label{eq:sigmaxzident}
  A^{-M}_M(\sigma_z^{(\N)};\qv)\equiv\bra \wfbra -M|\sigma_z^{(\N)}\,\phasefactor
  |\wf M\ket=
  \bra \wfbra M|\sigma_x^{(\N)}\,\phasefactor
  |\wf M\ket\equiv A^{M}_M(\sigma_x^{(\N)};\qv)
  \;\; .
\end{equation}
Ultimately then, there are five non-equal, non-zero ${}^3$He matrix elements
of the one-body operators
$\left\{\sigma_0^{(N)},\sigma_x^{(N)},\sigma_y^{(N)},\sigma_z^{(N)}\right\}$,
out of a possible $16$. This matches the five independent transition densities
after time-reversal, Hermitecity, and flipping symmetry have been applied.

The proof of the two-body relation~\eqref{eq:symmetry-twobody} proceeds
analogously to that of eq.~\eqref{eq:symmetry-onebody-general}. Since the
Compton two-body operator $\hat{O}_{12}$ is time-reversal even and the matrix
element is real below all thresholds, one finds by inserting
eqs.~\eqref{eq:Tsym} and \eqref{eq:Tsym-photon}:
\begin{equation}
  \label{eq:twobodysymmetry}
\begin{split}
  \bra \wfbra M'|\bra \lambda'| \mathcal{T} \hat{O}_{12} \mathcal{T}^{-1}
  |\lambda \ket|\wf M \ket&=\bra \wfbra- M'|\bra
  -\lambda'|\hat{O}_{12}|-\lambda \ket |\wf- M \ket\\
  & = \sign^{2J + M'+ M+ \lambda^\prime+ \lambda} \bra \wfbra M'|\bra \lambda'|
  \hat{O}_{12} |\lambda \ket|\wf M \ket 
\;\;. 
\end{split}
\end{equation}
Equation~\eqref{eq:symmetry-twobody} follows directly because $2(J+M+\lambda)$
is even for both half-integer and integer quantum numbers.

\end{document}